\documentstyle[12pt]{article}
\newcommand{\lbl}[1]{\label{eq:#1}}
\newcommand{\rf}[1]{(\ref{eq:#1})}
\newcommand{\be}{\begin{equation}}
\newcommand{\en}{\end{equation}}
\newcommand{\bea}{\begin{eqnarray}}
\newcommand{\ena}{\end{eqnarray}}
\newcommand{\np}[1]{Nucl.\ Phys.\ {\bf #1}}

\newcommand{\pr}[1]{Phys.\ Rev.\ {\bf #1}}
\newcommand{\prl}[1]{Phys.\ Rev.\ Lett.\ {\bf #1}}
\newcommand{\pl}[1]{Phys.\ Lett.\ {\bf #1}}
\newcommand{\ap}[1]{Ann. Phys. (NY)\ {\bf #1}}
\newcommand{\mpl}[1]{Mod.\ Phys.\ Lett.\ {\bf #1}}
\newcommand{\zp}[1]{Zeit. Phys.{\bf #1}}

\def\overleftrightarrow#1{\vbox{\ialign{##\crcr
$\leftrightarrow$\crcr\noalign{\kern-1pt\nointerlineskip}
$\hfil\displaystyle{#1}\hfil$\crcr}}}
\def\pip{{\pi^+}}

\def\piz{{\pi^0}}

\def\kz{{K^0}}
\def\kp{{K^+}}
\def\mkp{M_\kp}
\def\mkz{M_\kz}
\def\mv{M_V}
\def\mvp{M_{V'}}
\def\ma{M_A}
\def\mg{M_\gamma}
\def\muo{\mu_0}
\def\ms{m_s}
\def\mhat{ \hat{m} }
\def\fv{F_V}

\def\fac{ {1\over16\pi^2} }
\def\lag{{\cal L} }
\def\dslash{\partial\llap\slash}
\def\pslash{p\llap\slash}
\def\Dslash{D\llap{\llap\slash}}
\def\Aslash{A\llap\slash}
\def\Udag{{U^\dagger}}
\def\chidag{{\chi^\dagger}}
\def\qr{{q_R}}
\def\ql{{q_L}}
\def\F{F_\pi}
\def\intif{{\int_0^\infty}}
\def\mpi{M_\pi}
\def\mk{M_K}
\begin{document}
\rightline{IPNO/TH 97-03}
\rightline{(Revised version) June 1997}

\medskip
\begin{center}
{\large\bf A SUM RULE APPROACH TO THE VIOLATION OF DASHEN'S THEOREM }
\end{center}
\smallskip

\centerline{ B. Moussallam}

\centerline{\sl Division de 
Physique Th\'eorique\footnote{Unit\'e de recherche des
universit\'es Paris 11 et Paris 6 associ\'ee au CNRS} Institut de Physique
Nucl\'eaire}
\centerline{\sl Universit\'e Paris-Sud, 91406 Orsay France}
\vfill
\centerline{\large\bf Abstract}

A classic sum rule by Das et al. is extended 
to seven of the low-energy constants $K_i$, introduced by Urech,  
which parameterize electromagnetic
corrections at chiral order $O(e^2p^2)$. Using the spurion formalism, a
simple convolution representation is shown to hold
and the structure in terms of the
chiral renormalization scale, the QCD renormalization scale and the QED gauge
parameter is displayed. The role of the resonances is studied as providing
rational interpolants to relevant 
QCD n-point functions in the euclidian domain. 
A variety of asymptotic constraints must be implemented which have
phenomenological consequences.  
A current assumption concerning the dominance of the lowest-lying
resonances is shown clearly to fail in some cases.
\vfill
\eject
\leftline{}

\noindent{\large\bf 1. Introduction:}
\medskip

Thirty years ago, Das et al.\cite{das1} (DGMLY) derived a remarkable relation
between the mass difference of the charged and neutral pions and the
masses of the lightest vector and axial vector resonances,
\begin{equation}
M^2_\pip-M^2_\piz={3 e^2\over16\pi^2}{\ma^2\mv^2\over\ma^2-\mv^2}
                \log{\ma^2\over\mv^2} \ .		 
\lbl{dmpi}
\end{equation}                                    
This relation follows from a sum rule which is exact in the chiral
limit (i.e. $m_u=m_d=m_s=0$) under the only extra assumption that the
lowest-lying vector and axial vector meson resonances make the essential
contribution to the integral. 
The physical $\pip-\piz$ mass difference 
is nearly purely electromagnetic in origin and
happens to be rather accurately
described by eq.\rf{dmpi}. The analogous mass difference of kaons,
$M^2_\kp-M^2_\kz$, has an electromagnetic contribution and a purely QCD
contribution proportional to $m_u-m_d$ which are approximately of the same
magnitude. Knowledge of the electromagnetic contribution allows one to
access the value of the quark mass difference $m_u-m_d$ (divided by, say,
$m_u+m_d$) using chiral perturbation theory\cite{gl85}. 
It has long been believed that estimating the EM
contribution to $M^2_\kp-M^2_\kz$ in the chiral limit was sufficient. 
In this limit, it is
given by Dashen's theorem\cite{dashen} (DT) to be equal to $M^2_\pip-M^2_\piz$.
This approximation is now known to fail for the purpose of extracting
$m_u-m_d$. In particular, the value of the $\eta$ decay rate
$\Gamma(\eta\to3\pi)$ that one would predict (at one loop and including
estimates of higher loop corrections \cite{kambor}\cite{anisovich}) would be
too small by as much as a factor of two compared to experiment. 

Thus, it is necessary to estimate the EM contributions to $M^2_\kp-M^2_\kz$
beyond the chiral limit (the difference from $M^2_\pip-M^2_\piz$ is 
customarily, but
somewhat inappropriately, referred to as the violation of Dashen's theorem). 
There have been many  
attempts over the years in this direction\cite{socolow}\cite{pagels}
\cite{maltman}\cite{dhw}\cite{bij1}\cite{b+u}\cite{bij2}
(a representative, but non-exhaustive list)
and, very recently, a lattice
calculation has appeared\cite{lattice}. 
Several rather different approaches to this problem have been followed. 
The work of refs.\cite{dhw} and
\cite{b+u} is based on the assumption that most of the information 
on DT violation  is contained in the parameters 
of the low lying resonances, in a way  similar to eq.\rf{dmpi}. 
Unfortunately, the results of these two papers (which
should be identical) are not in very good
agreement which each other\footnote{
A diagram was
incorrectly evaluated in ref.\cite{dhw}. An update of this calculation has
appeared very recently\cite{perez}, 
which obtains a very similar result numerically. }.  
One of the purposes of the present paper is to investigate in some detail
the validity of this assumption. 

As shown by Urech\cite{urech}, at low energies, 
electromagnetic effects can be
parametrized in a chiral lagrangian framework
\cite{weinberg}\cite{gl84}\cite{gl85}. 
In this formalism, the leading electromagnetic
contributions are computed from the effective lagrangian at tree level,
which contains a single low-energy constant (LEC): the result of Das et al.
provides a sum rule for this LEC. At next-to-leading order, the
electromagnetic corrections are obtained by computing the photon loop 
(as well as the pion loops) from
the leading order lagrangian 
and adding the contributions at tree level from the next order terms in the 
lagrangian, which involves 
(essentially\footnote{There are further constants which either play no role
at low energy or correspond to $e^4$ contributions.}
) 13 new low-energy constants  $K_1, K_2,...,K_{13}$. 
In this paper, we will propose a  sum rule evaluation of 
the seven parameters $K_7,...,K_{13}$. 
The $\kp-\kz$ mass difference actually 
involves two more constants, $K_5$ and $K_6$, 
which we will not attempt to evaluate. 
Little was known, up to now, on the individual values of these LEC's
(estimates for some combinations were given recently\cite{bij2})
except for their order of magnitude which, for consistency of the chiral
counting for the electric charge, should be the same as that of the
ordinary $O(p^4)$ constants $L_i$.  
Individual knowledge of these constants is useful for the purpose of 
evaluating radiative corrections. Indeed, as precision
increases in the predictions of chiral perturbation theory, the computation
of radiative corrections is becoming a topic of increased interest. This is
already true for the $\pi\pi$  scattering amplitude 
which is now known at the level of
two loops\cite{orsay}\cite{berne}, and which will be subjected to a
precision test in the planned  experiment DIRAC\cite{dirac}.

In general, we will show that the $K_i$'s can be expressed as a convolution
of a QCD correlation function with the electromagnetic propagator, plus
a contribution from the QED counterterms which remove the divergence of
the integral. The LEC's  
$K_1,...,K_6$ are related to QCD 4-point functions, while
$K_7,...,K_{13}$ are related to QCD two- and three-point functions. In that
case, we will show that the contribution of the resonances can be
discussed independently of any specific lagrangian model for resonances. 
An important question concerns 
the validity of the approximation of retaining only the lightest 
multiplet of resonances in each channel.
One may view resonance saturation as a method of constructing rational
interpolants to the QCD n-point functions in the euclidian region. 
These interpolants must satisfy
certain asymptotic constraints, in order for the QED divergencies to cancel
out, and one also expects them to be reasonably precise at low momenta, i.e.
to some extent in the resonance region as well. It is not clear that low
order interpolants are capable of satisfying all these constraints.

The plan of the paper is as follows. In the next section, we discuss
the form of the QED counterterms when  the spurion formalism 
is applied to the 
electric charge. The use of this formalism is a key technical ingredient
in the derivation of the sum rules. In sec.3, we discuss a set of sum rules
for the parameters $K_{11}$, $K_{12}$ and $K_{13}$ which involve QCD
correlators in the chiral limit. We will show that a single 
multiplet of vector and of axial-vector resonances is enough to 
obey adequately 
all the asymptotic constraints, and we obtain expressions which are
rather neat generalizations of eq.\rf{dmpi}. 
Next, in sec.4, we discuss $K_7$, $K_8$, $K_9$ and $K_{10}$. 
The first two are suppressed by the Zweig rule and the latter two 
can be expressed
in terms of flavour symmetry breaking differences of vector and axial-vector
spectral functions. The phenomenology in terms of resonances is then discussed
in sec.5 and the application to the violation of Dashen's theorem in sec.6.

\noindent{\large\bf 2. The spurion formalism and the QED counterterms}

The spurion formalism applied to the electric 
charge matrix\cite{egpr} $q={\rm diag}(2/3,-1/3,-1/3)$ consists
in calling $q$ by two different names $q_L$ and $q_R$ such that the
QED interaction lagrangian for the quarks can be split into two pieces:
\be\lbl{lqed1}
\lag_{QED}=i\bar\psi(\dslash+iq\Aslash)\psi\equiv
i\bar\psi_L(\dslash+iq_L\Aslash)\psi_L+
i\bar\psi_R(\dslash+iq_R\Aslash)\psi_R
\en
(with $\psi^t=(u,d,s)$ ). 
One may thus render the QED lagrangian invariant under the chiral group
by assuming appropriate chiral transformation properties for the two
spurions $q_L$ and $q_R$. This is the same method as has been applied to
the mass matrix\cite{gl84}. In order to take quantum electrodynamical
effects into account, the chiral lagrangian must include the photon as a
dynamical field, in addition to the octet of pseudo-Goldstone boson fields. As
usual, the chiral lagrangian will consist of the most general set of local
interaction terms which are invariant under the chiral group and which are
classified according to increasing chiral order. In the EM sector, 
the correct counting rules, introduced by 
Urech\cite{urech}, state that the photon
field $A_\mu$ is of order $O(p^0)$ and the charge spurions $q_L$ and $q_R$
are of order $O(p)$. 
At order two, the chiral lagrangian consists, firstly, of the ususal
terms,
\be\lbl{lag2a}
{\cal L}^{(2)}_a={F^2_0\over4}<D_\mu U D^\mu U^\dagger +U \chi^\dagger +\chi
U^\dagger>,
\en
where, however, the covariant derivatives contain $q_L$ and $q_R$ in
addition to the usual vector and axial-vector external sources
\be\lbl{dmuU}
D_\mu U=\partial_\mu U-i(v_\mu+a_\mu+q_R A_\mu) U +iU(v_\mu -a_\mu +q_L
A_\mu). 
\en
In addition to that, one must include the purely photonic lagrangian, 
\be\lbl{lag2b}
{\cal L}^{(2)}_b=-{1\over4}F_{\mu\nu}F^{\mu\nu},
\en
and, finally, there is a single invariant term
with two spurions that one can form\cite{egpr},
\be\lbl{lag2c}
{\cal L}^{(2)}_c=C<q_R U q_L U^\dagger>\ .
\en
Equations \rf{lag2a}, \rf{lag2b}, \rf{lag2c} 
define the most general lagrangian of chiral order two which 
satsisfies EM gauge invariance in the limit of constant spurions. 
In order to quantize the electromagnetic field, one must include a gauge
fixing term and it is also useful to introduce a small photon mass
$M_\gamma$ in order to regulate infrared divergencies at
intermediate steps of the calculations, this is performed by adding
\be
{\cal L}^{(2)}_d=
{1\over2}M^2_\gamma\, A_\mu
A^\mu- {1\over2\xi}\,(\partial_\mu A^\mu)^2\ .
\en
The corresponding photon propagator is
\be\lbl{prop}
-iD_{\mu\nu}(p)={-i\over p^2-\mg^2}\left[
g_{\mu\nu}+(\xi-1){p_\mu p_\nu\over p^2-\xi\mg^2}\right]\ .
\en
At chiral order two, Green's
functions and observables are computed by using ${\cal L}^{(2)}$ at tree
level. At the next chiral order, $O(p^4)$, one must firstly 
compute the one-loop graphs with vertices  generated from 
${\cal L}^{(2)}$. The loops  will involve both
pion and photon internal lines in the present case. Secondly, one must
add the tree contributions
from ${\cal L}^{(4)}$. 
The lagrangian ${\cal L}^{(4)}$ will consist, firstly, of
the usual terms of Gasser and Leutwyler.\cite{gl85}. In addition, one can
construct a set of terms with two EM spurions
\footnote{Symmetry considerations alone do not rule out 
terms with a single spurion and no $A_\mu$, for example a term like 
$<\ql D_\mu U^\dagger
D^\mu U +\qr D_\mu U D^\mu U^\dagger>$. The technique to be used repreatedly
below immediately shows that the coefficients of such terms vanish because
the vacuum expectation value of the photon field does.}, 
which were classified by
Urech. For convenience we will separate the terms which involve two
derivatives of the pion fields, 
\bea\lbl{lag41-6}
\lag^{(4)}_{1-6}
&&={1\over2}K_1 F_0^2<D_\mu U D^\mu U^\dagger><\ql\ql+\qr\qr>\\ \nonumber
&&+K_2 F_0^2<D_\mu U D^\mu U^\dagger><\ql U^\dagger \qr U>\\ \nonumber
&&+K_3 F_0^2(<D_\mu U\ql U^\dagger>^2 +<D_\mu U^\dagger\qr U>^2)\\ \nonumber
&&+K_4 F_0^2 <D_\mu U\ql U^\dagger><D_\mu U^\dagger\qr U>\\ \nonumber
&&+K_5 F_0^2<D_\mu U^\dagger D^\mu U\ql\ql+ D_\mu U D^\mu U^\dagger\qr\qr>\\
\nonumber
&&+K_6 F_0^2<D_\mu U^\dagger D^\mu U\ql U^\dagger \qr U +D_\mu U
D^\mu U^\dagger\qr U \ql U^\dagger>, 
\ena
from the remaining ones
\bea\lbl{lag4}
\lag^{(4)}_{7-13}
&&= {1\over2}K_7 F_0^2<\chidag U +\Udag\chi><\ql\ql+\qr\qr> \\ \nonumber
&&+          K_8 F_0^2<\chidag U +\Udag\chi><\ql\Udag\qr U>\\ \nonumber
&&+K_9 F_0^2<(\chidag U +\Udag\chi)\ql\ql +
             (\chi   \Udag+ U\chidag)\qr\qr >\\ \nonumber
&&+K_{10} F_0^2<(\chidag U   +\Udag\chi)\ql\Udag\qr U +
                (\chi   \Udag+ U\chidag)\qr U\ql\Udag>\\ \nonumber
&&+K_{11} F_0^2<(\chidag U -\Udag\chi)\ql\Udag\qr U +
                (\chi   \Udag- U\chidag )\qr U\ql\Udag>\\ \nonumber
&&+K_{12} F_0^2<U D^\mu\Udag[D_\mu\qr,\qr] +
             \Udag D^\mu U[D_\mu\ql,\ql]>\\ \nonumber
&&+K_{13} F_0^2<D_\mu\qr\, U D^\mu\ql\,\Udag>,
\ena
where
\be
D_\mu q_L=\partial_\mu q_L-i[v_\mu-a_\mu,q_L],\quad
D_\mu q_R=\partial_\mu q_R-i[v_\mu+a_\mu,q_R]\ .
\en
Further terms which contain four EM spurions as well as terms which vanish
when $A_\mu$ is set to zero  will not be considered in the following. 

All these terms are counted as $O(p^4)$ according to the
standard chiral counting rules\cite{gl84}. According to the so-called
generalized CHPT (see e.g. ref.\cite{daphne})  
quark masses are counted as $O(p)$ rather than $O(p^2)$ such that 
the terms corresponding to
$K_{7}$ up to $K_{11}$ would be counted as of chiral order three. 
In this scheme, 
the terms corresponding to $K_1$ to $K_6$ as well as $K_{12}$, $K_{13}$ 
would appear at $O(p^4)$ as before but, to the same order,
a number of additional terms with two occurences of the scalar 
field $\chi$  must also be included. 
This modification of the chiral counting corresponds to a weak
form of spontaneous chiral symmetry breaking in which $F_\pi\ne0$ while the 
quark condensate
$<\bar q q>$ could be small or even vanishing. Whether 
this scheme is of physical relevance or not can be decided experimentally 
by studying the low-energy $\pi\pi$ amplitude, which is very sensitive to
the size of the quark condensate \cite{daphne}\cite{orsay}. 
In this paper, the discussion will be restricted to the case of
the standard chiral counting, which is the most predictive framework. 

The effective lagrangian generates a low-energy representation for the
generating functional of QCD Green's functions $W$. In addition to the sources
$s(x),\ p(x),\ v_\mu(x),\ a_\mu(x)$ that one customarily introduces, $W$ now
depends on the two new sources $q_L$ and $q_R$,
\be
W\equiv W(v_\mu,a_\mu,s,p,q_L,q_R)\ .
\en
Accordingly, one can introduce Q-currents,
\be\lbl{qcurrent}
Q^a_L={\delta W\over i\delta q^a_L},\quad
Q^a_R={\delta W\over i\delta q^a_R}\ .
\en
and one may identify the low-energy constants by taking functional
derivatives of $W$ 
with respect to $q_L$ and $q_R$. It is useful, in connection with
the LEC's $K_{12}$ and $K_{13}$, 
to further extend this formalism by letting $q_L$ and $q_R$
become space-time dependent. This generalization allows one to 
identify $K_{13}$, for
instance, from the two-point function $\delta^2
W/\delta q_L(x)\delta q_R(0)$ instead of a four-point function if the
spurions are restricted to be constants. 
There is 
an apparent drawback to this extension, which is 
that we are allowing the photon to
couple to a nonconserved current.  This is an unusual situation 
and such a theory  is not guaranteed to be renormalizable. 
It is not clear whether it is possible to define the generating functional
in full generality in the presence of $x$ dependent spurions. For our
restricted purposes (i.e. the derivation of sum rules which are UV
finite for $K_{12}$ and $K_{13}$) we simply need the expansion of $W$ to
quadratic order in $q_L$, $q_R$ and we can switch off the external
electromagnetic field. In this case, the UV divergencies are removed by the
following finite set of counterterms
\be\lbl{lspur2}
\lag_{q^2}={i\over2}Z_2\bar\psi_L 
q_L\overleftrightarrow{\Dslash}q_L\psi_L +(L\leftrightarrow R)
               -  Z_s\bar\psi_R q_R(s+i p) q_L \psi_L
               -  Z_s\bar\psi_L q_L(s-i p) q_R \psi_R
\en
The existence of such a set of local counterterms guarantees the validity
of the idea of using space-time dependent spurions at this order. 
For our purposes, \rf{lspur2} will be used at tree level. We may 
then use the equations of motion generated by \rf{lqed1}, 
thereby obtaining the counterterms in the  form
\bea\lbl{lspur2a}
\lag_{q^2}=&&{i\over2}
Z_2 \bar\psi_L  [q_L,D^\mu q_L]\gamma_\mu\psi_L 
+(L\leftrightarrow R) \\ \nonumber
&&
+ \bar\psi_R\left[ {1\over2}{Z}_2\left( q^2_R(s+i p)+(s+i p)q^2_L\right)
                          -{Z}_s q_R(s+i p)q_L \right]\psi_L\\ \nonumber
&&
+\bar\psi_L\left[ {1\over2}{Z}_2\left( q^2_R(s-i p)+(s-i p)q^2_L\right)
-                         {Z}_s q_L(s-i p)q_R \right]\psi_R\ . 
\ena
Setting $q_L=q_R=q$, one recovers the usual QED counterterms.
For a free fermion, the renormalization constants $Z_2$ and $Z_s$ are
determined from the requirement 
that the fermion self energy $\Sigma(\pslash)$ is 
normalized in the way appropriate for a free particle of mass $m$. 
For instance, in dimensional regularization, $Z_2$ and $Z_s$ read
\bea\lbl{ZZ}
&&Z_2(m)={\muo^{-2\epsilon}\over16\pi^2}\left\{
-\xi\Big(\Gamma(\epsilon)+1-\log{m^2\over4\pi\muo^2}\Big)
+(\xi-3)\log{\mg^2\over m^2}+\xi\log\xi-3\right\}\\ \nonumber
&&Z_s(m)={\muo^{-2\epsilon}\over16\pi^2}\left\{
-(\xi+3)\Big(\Gamma(\epsilon)+1-\log{m^2\over4\pi\muo^2}\Big)
+(\xi-3)\log{\mg^2\over m^2}+\xi\log\xi+2\right\} ,
\ena
Quarks, however, are not free particles but on the contrary, 
are confined, so it seems unwise to use the above values of $Z_2$ and 
$Z_s$ in the present context. 
In fact, the QED counterterms 
play a role in a kinematical region where QCD becomes perturbative. 
The most reasonable thing to do, then, seems to 
resort to $\overline{MS}$ renormalization, as one  
customarily does in perturbative QCD calculations. Such a mass independent
renormalization scheme is in fact implicitly assumed in the definition 
of the \lq\lq current" quark masses, which appear in the effective
lagrangian (see e.g. the review by Manohar in the PDG\cite{pdg96}).  
As a consequence, 
the corresponding renormalization scale $\mu_0$ will appear
in the expression of some of 
the low-energy constants. This was pointed out in ref.\cite{bij2}. 
According to this prescription, $Z_2$ and $Z_s$ have the following expression, 
\bea\lbl{ZZb}
&&Z_2={\muo^{-2\epsilon}\over16\pi^2}
(-\xi)\Big[\Gamma(\epsilon)+\log(4\pi) \Big]\\ \nonumber
&&Z_s={\muo^{-2\epsilon}\over16\pi^2}
(-\xi-3)\Big[\Gamma(\epsilon)+\log(4\pi)\Big]\ .
\ena
which will be adopted in the following.
Further divergences involving the strong coupling constant $\alpha_s$ will
be assumed to be removed by the $\overline{MS}$ prescription as well.

\newpage
\noindent{\large\bf 3. QCD in the chiral limit: 
sum rules for $C$, $K_{11}$, $K_{12}$ and $K_{13}$}

\noindent{\bf 3.1 The constants $C$ and $K_{13}$: spectral representation }

Let us consider the correlation function $<Q_L Q_R>$ constructed from 
the currents associated with the spurion fields $\ql$ and $\qr$ introduced
in sec.2,
\bea
\Pi^Q_{LR}(p^2)&&=4i\int d^4x {\rm e}^{ipx}<0\vert T Q^3_L(x) Q^3_R(0)\vert0>
\\ \nonumber
&& =
i\int d^4x {\rm e}^{ipx}<0\vert T Q^3_V(x) Q^3_V(0)-Q^3_A(x) Q^3_A(0)\vert0>
\ena
We have introduced vector and axial vector Q-currents:
\be
Q_V=Q_L+Q_R\qquad Q_A=-Q_L+Q_R\ .
\en

On the one hand, we can use the chiral lagrangian, and compute the chiral
expansion of $\Pi^Q_{LR}$ up to chiral order four. 
It receives contributions at
tree level involving the constants $C$ and $K_{13}$ and a photon loop
contribution. There is no pion loop contribution. Explicitly, a simple 
calculation gives, in an arbitrary gauge\footnote{
We follow the usual chiral $\overline{MS}$ 
convention of relating bare and renormalized constants
by: $K_i=\mu^{-2\epsilon}[K^r_i(\mu)-\Sigma_i/32\pi^2
(\Gamma(\epsilon)+1+\log(4\pi))]$}
\be\lbl{lrdv}
\Pi^Q_{LR}(p^2)=2C+p^2 F_0^2\left\{ 2 K^r_{13}(\mu)+\fac{3\over4}
\bigg[ (\xi-1)\Big(\log{\mg^2\over\mu^2}+{1\over6}\Big)
+\xi\log\xi\bigg]+O(p^2)\right\}\ .
\en
On the other hand, it is clear from \rf{lqed1} that each Q-current is nothing
but a QCD current multiplied by the quantum photon field, so that the above
correlator can be written as a convolution of a QCD correlation function
with the photon propagator:
\be\lbl{lrconv}
\Pi^Q_{LR}(p^2)=\int{d^4k\over(2\pi)^4} \left(\Pi^{\alpha\beta}_{V^3}(k-p)
-\Pi^{\alpha\beta}_{A^3}(k-p)\right)
\left(-iD_{\alpha\beta}(k)\right)\ .
\en
In this case, the QED counterterms (see \rf{lspur2a})
make no contribution, implying that the integral must be finite. 
This can be verified by using the operator product expansion for 
the difference $\Pi^{\alpha\beta}_{V^3}-\Pi^{\alpha\beta}_{A^3}$ 
which, indeed, starts with dimension six operators in the chiral limit. 
The vector and axial-vector currents are defined, as usual, as
\be\lbl{currents}
V^i_\mu(x)=\bar\psi(x)\gamma_\mu{\lambda^i\over2}\psi(x), \qquad
A^i_\mu(x)=\bar\psi(x)\gamma_\mu\gamma^5{\lambda^i\over2}\psi(x)\ .
\en 
The $<VV>$ and $<AA>$
correlation functions which appear in \rf{lrconv}
satisfy once-subtracted  dispersive representations (which we display in
unsubtracted form for simplicity and for the reason that we will only
consider combinations which need no subtractions in the sequel). 
Assuming the T-product to be 
covariantly defined, 
\be\lbl{specv}
\Pi^{\mu\nu}_{V^3}(q)=i\int d^4x\, {\rm e}^{iqx}<0\vert T\, V^{3\mu}(x)
V^{3\nu}(0)\vert0>
=(q^\mu q^\nu-g^{\mu\nu}q^2){1\over\pi}\int_0^\infty{ds\over s-q^2}\,
\rho_{V^3}(s)
\en
and
\bea\lbl{speca}
\Pi^{\mu\nu}_{A^3}(q)&&=i\int d^4x\, {\rm e}^{iqx}<0\vert T\,A^{3\mu}(x)
A^{3\nu}(0)\vert0>\\ \nonumber
&&=(q^\mu q^\nu-g^{\mu\nu}q^2){1\over\pi}\int_0^\infty{ds\over s-q^2}
\,\rho_{A^3}(s)+g^{\mu\nu}{1\over\pi}\int_0^\infty{ds\over s-q^2}
\,\sigma_{A^3}(s)\ .
\ena
In the chiral limit, the spectral function $\sigma_{A^3}(s)$ vanishes
identically. It is convenient to separate explicitly the pion contribution
in the axial spectral function,
\be
\rho_{A^3}(s)=\bar\rho_{A^3}(s)+F_0^2\delta(s)\ .
\en
It is now a simple matter to insert the spectral representations 
\rf{specv}\rf{speca} into the expression \rf{lrconv} for $\Pi^Q_{LR}$ 
and compute the photon loop integral. 
Comparing with the chiral expansion form of $\Pi^Q_{LR}$ \rf{lrdv}, 
one readily
identifies the constants $C$ and $K_{13}$ for which one obtains the
following representations:
\be\lbl{C}
C=-\fac{3\over2\pi}\intif dx\, x \log{x\over\mu^2}\big(\rho_{V^3}(x)
-\bar\rho_{A^3}(x) \big),
\en
which is a re-derivation of the DGMLY sum rule\cite{das1} 
(using the relation $M^2_\pip-M^2_\piz=2e^2C/F^2_0$ which is exact in the
chiral limit and to order $O(e^2)$), 
and its immediate generalization,
\be\lbl{K13}
K^r_{13}(\mu)=\fac{3\over4}\left\{1 +(1-\xi)\bigg[{1\over12}+
{1\over2\pi F_0^2}\intif dx \log{x\over\mu^2}(\rho_{V^3}(x)-\bar\rho_{A^3}(x))
\bigg]\right\}
\ .
\en
The expected order of magnitude for the constants 
$K^r_i(\mu=M_V)\simeq1/16\pi^2$ is indeed
confirmed by this explicit calculation, but one must be careful since
these constants are gauge dependent in general.  
The dependence of $C$ and $K_{13}$ upon the chiral renormalization scale
$\mu$ can be deduced from eq.\rf{lrdv} since the left-hand side is
independent of $\mu$: C is a constant and $K_{13}$ satisfies
\be
\mu\,{dK^r_{13}(\mu)\over d\mu}={1\over16\pi^2}{3\over4}(\xi-1)\ .
\en
The correct scale dependence emerges from the spectral representations
\rf{C} and \rf{K13} provided the spectral functions satisfy the two Weinberg
sum rules\cite{wrules} (a modern discussion can be found in ref.\cite{wbook})
\be\lbl{wrules}
{1\over\pi}\intif dx\, (\rho_{V^3}(x)-\bar\rho_{A^3}(x))=F_0^2,\quad
\intif dx\, x\, (\rho_{V^3}(x)-\bar\rho_{A^3}(x))=0\ .
\en

\noindent{\bf 3.2 The LEC's $K_{11}$ and $K_{12}$ and the correlator $<VAP>$}

We will follow the same approach as in sec. 3.1. In order to isolate 
contributions from the LEC's $K_{11}$ and $K_{12}$ we may consider a
correlation function of two Q-currents with the pseudoscalar current,
\be\lbl{piw}
\Pi^Q_W(p,q)=\int d^4x d^4y\, {\rm e}^{ipx+iqy}
<0\vert T Q^1_V(x) Q^2_A(y) P^3(0)\vert 0>\ ,
\en
where the pseudoscalar current is defined as 
\be
P^a(x)=i\bar\psi(x)\gamma^5{\lambda^a\over2}\psi(x)\ .
\en
It will also be useful to introduce the related pion to vacuum 
matrix element,
\be\lbl{piwh}
\Pi^Q_{\hat W}(p,l)=\int d^4x\, {\rm e}^{ipx}
<0\vert T Q^1_V(x) Q^2_A(y) \vert \pi^3(l)>\ .
\en
We first construct the chiral expansions of these objects, up to one loop, 
using the chiral lagrangian. Setting $q=0$ in \rf{piw}, we obtain
\be\lbl{2k11-k12}
\Pi^Q_W(p,0)={-2B_0 C\over p^2}-2B_0 F_0^2\, (2K_{11}-K_{12}) + O(p^2)\ .
\en
The contributions from the photon loop and from the pion loop
both vanish, implying that the combination $2K_{11}-K_{12}$ is finite in any
gauge. Setting $p=0$ now in \rf{piw}, we obtain the following chiral
expansion, which involves the combination $2K_{11}+K_{12}$,
\bea\lbl{2k11+k12}
\Pi^Q_W(0,q)&&={-2B_0 C\over q^2}\\ \nonumber
&&-B_0 F_0^2\left\{
2\Big(2K^r_{11}(\mu)+K^r_{12}(\mu)\Big)
+\fac\left[ (\xi-{3\over2})\log{\mg^2\over\mu^2}+\xi\log\xi-{1\over4}\right]
+O(q^2)\right\}\ .
\ena
We will also use the expansion of $\Pi^Q_{\hat W}$:
\bea\lbl{2k12+k13}
\Pi^Q_{\hat W}(p,l)&&=i{2C\over F_0}+iF_0, p\cdot l\,\bigg\{
-4K^r_{12}(\mu)-2 K^r_{13}(\mu)\\ \nonumber
&&+\fac{1\over4}\bigg[
(-7\xi+9)\log{\mg^2\over\mu^2}-7\xi\log\xi-{1\over2}\xi+{3\over2}\bigg]
+O(p^2,p\cdot l)\bigg\} + O(p^4)\ .
\ena
We may now express $\Pi^Q_W$ and $\Pi^Q_{\hat W}$ by functionally 
differentiating the QCD generating functional  with respect to the sources
$q_L(x)$, $q_R(x)$ and $p(x)$, and one obtains,
\be\lbl{wint}
\Pi^Q_W(p,q)=
\int {d^4k\over (2\pi)^4} W_{\mu\nu}(p-k,q+k)
(-i D^{\mu\nu}(k) )+{1\over2}B_0 F^2_0\left[ Z_2 +Z_s 
-2Z_2{p.(p+q)\over (p+q)^2}\right]\ ,
\en
where $W^{\mu\nu}(p,q)$ is the correlator of one vector, one axial-vector
and one pseudoscalar current $<VAP>$ in the chiral limit,
\be\lbl{w}
W_{\mu\nu}(p,q)=\int d^4x d^4y\, {\rm e}^{ipx+iqy}
<0\vert T V^1_\mu(x) A^2_\nu(y) P^3(0)\vert0>\ .
\en
In eq.\rf{wint} the convolution integral simply arises from replacing the
currents $Q_V$ and $Q_A$ from their definitions. The  additional contribution 
gets
generated upon functionally differentiating the QED counterterms. This
contribution implies that the convolution integral is UV divergent in QCD,
in such a way that the sum of the two terms in eq.\rf{wint} is finite.

Equating expression \rf{wint} with the chiral expansions \rf{2k11-k12} and
\rf{2k11+k12} one obtains an expression for the LEC's $K_{11}$ and $K_{12}$
as a convolution of the QCD 3-point function $<VAP>$ and the electromagnetic
propagator. An additional expression, which we have used to cross-check the
results,  can be found which involves the pion to
vacuum matrix element $<0\vert VA\vert\pi>$, 
\be
\hat W_{\mu\nu}(p,l)=\int d^4x {\rm e}^{ipx}<0\vert T V^1_\mu(x)
A^2_\nu(0)\vert \pi^3(l)>={-i\over B_0 F_0}
\lim_{l^2\to 0}l^2 W_{\mu\nu}(p,q)\ .
\en
The Q-current matrix element introduced in eq.\rf{piwh} can be expressed 
in terms of $<0\vert VA\vert\pi>$ as 
\be\lbl{whint}
\Pi^Q_{\hat W}(p,l)=\int{d^4k\over (2\pi)^4} {\hat W}_{\mu\nu}
(p-k,l)(-i D^{\mu\nu}(k) )
+iF_0 Z_2\, p\cdot l\ , 
\en
and using the chiral expansion \rf{2k12+k13} one obtains a convolution
representation for the combination $2K_{12}+K_{13}$.

To summarize, at this point we have shown that the constants $K_{11}$,
$K_{12}$ and $K_{13}$ can be expressed as convolution representations
in terms of QCD correlators in the chiral limit. 
Using the same ideas it is not difficult 
to see that the constants $K_1$ to $K_6$ must
obey similar representations, involving the 
QCD correlators $<VVAA>$ and $<AAAA>$ in the chiral limit as well. The
difficulty in this case is the implementation of resonance saturation. 
We will now show that a rather neat implementation can be performed for the
correlators $VV-AA$ (which is very simple) and $<VAP>$ ( which is
somewhat less trivial) and defer a brief discussion of $K_1$ to $K_6$ to sec. 
3.6. 

\noindent{\bf 3.3 Resonance saturation of $<VV-AA>$}

The spectral densities $\rho_V$ and $\rho_A$ which appear in the expressions
\rf{C} and \rf{K13} for $C$ and $K_{13}$ can be related to exprimentally
measurable quantities like $e^+e^-\to hadrons$ cross sections and $\tau$
decay rates. Analyses of the DGMLY and Weinberg sum rules in terms of
available experimental 
data have been performed \cite{peccei}\cite{d+g}. A physically
appealling and time honoured approximation is that of \lq\lq resonance
saturation" of the spectral integrals. 
There are, in fact, two aspects in this approximation. One is to represent the
spectral functions as a sum over delta functions, and the second is to retain
just a few terms in the sum. The first aspect corresponds to a leading large
$N_c$ limit and was actually shown to be rather precise  in
practice \cite{d+g}.
However, even in the large $N_c$ limit it is not clear how many
resonances should be included. A current
assumption is that minimal saturation, i.e. in the present case including
only the contribution of the $\rho(770)$ and the $a_1(1260)$ 
resonances, does
provide a resonable approximation. In this case, the spectral
functions are given by
\be
\rho_{V^3}(x)=F^2_V\,\delta(x-M^2_V),\quad
\bar\rho_{A^3}(x)=F^2_A\,\delta(x-M^2_A)\ .
\en
One may at least verify that this lowest
order approximation is not grossly inconsistent. Indeed, if one uses
experimental values for $F_V$ and $M_V$ then the Weinberg sum rules are
satisfied with values of $F_A$ and $M_A$ which are within 30\% of those
of experiment. Furthermore, the corresponding result for the constant
$C$ seems to be rather accurate as one can judge from the resulting value of
$M^2_\pip-M^2_\piz$. In the dispersive representation, one may attribute the
success of the minimal resonance approximation to the fact that the various
integrands are oscillatory. In that situation, numerical analysis teaches us
that an optimal sequence of approximations is obtained by cutting the
integrand after an even number of oscillations. The precision could be much
better than one could expect from the asymptotic behaviour of the absolute
value of the integrand. This also suggests that in order to improve on the
first order approximation one should include an additional pair of resonances
$\rho'$ and $a_1'$, which means integrating up to a rather high
energy.

An alternative point of view on resonance saturation emerges from 
considering the correlation function $<VV-AA>$ itself. Resonance
saturation amounts to constructing rational
parametrizations for the correlation function. In the minimal version, for
instance, one would have
\be\lbl{ratva}
\Pi^{\mu\nu}_{V^3}(p)-\Pi^{\mu\nu}_{A^3}(p)=(p^\mu p^\nu-g^{\mu\nu}p^2)
\left\{{F^2_0\over p^2}+{F^2_A\over  p^2-M^2_A}-{F^2_V\over  p^2-M^2_V}
+polynomial\right\}\ .
\en
This is the most general form which satisfies the chiral Ward identities
(which dictate the form of the tensor structure) and which has a specified
number of resonance poles. 
When computing the four-dimensional
convolution integrals for $C$ or $K_{13}$, one may perform a
Wick rotation such that the correlator $<VV-AA>$ is needed only for
euclidian values of $p^2$. In this region, the correlator is a smooth 
function so it makes perfectly good sense to employ a rational
approximation. The parameters of the rational function \rf{ratva} 
($M_V$, $M_A$, $F_V$,...)  
may be constrained both from the asymptotic
region $p^2\to-\infty$, using information from the operator product expansion,
and by extrapolating
to positive values of $p^2$ using physical information on the
resonances. In the asymptotic region, 
the behaviour must ensure that the convolution integral 
defining $\Pi^Q_{LR}$ eq.\rf{lrconv} 
is finite. This implies that the polynomial in eq.\rf{ratva} must vanish 
and one must impose the two
conditions that the coefficients of $1/p^2$  and $1/p^4$ both vanish.  
These conditions are exactly equivalent to the
two Weinberg sum rules. 
We recover here the necessity of imposing asymptotic
matching conditions when discussing resonances in the effective lagrangian
context, a fact which has been appreciated relatively 
recently \cite{eglpr} (see
also \cite{revecker}). The usefulness of the effective
lagrangian for the resonances is to guarantee that  the n-point functions
that one might compute  automatically obey all the  chiral Ward
identities. When discussing two- or three-point functions it is sometimes more
expedient to write directly the most general form compatible with a given
set of resonance poles and impose that the relevant Ward identities be
obeyed by hand. This is what we have done for $<VV-AA>$ and we will proceed
in this way also for $<VAP>$ below.

Returning to the evaluation of the low-energy constants, after imposing the
two asymptotic conditions on the parametrization \rf{ratva}, 
performing the Wick rotation and integrating over
angles one obtains $C$ in the form of an integral
\be
C={3F^2_0M^2_AM^2_V\over 32\pi^2}\int_0^\infty{dx\over (x+M^2_V)(x+M^2_A)}\ .
\en
The interest of this expression is the observation that a substantial part of
the integral (approximately 40\%) comes from the high energy region 
$x>1\ {\rm GeV}^2$. This means that the first nonvanishing term in the
asymptotic expansion of $<VV-AA>$, call it $c/p^6$, plays an important role.
In the minimal saturation approximation $c$ is predicted to be
$c=F^2_0M^2_VM^2_A$. In QCD, $c$ is not exactly a constant because of the
anomalous dimension of the dimension six operator involved. Ignoring this
fact, and using the vacuum saturation approximation\footnote{
This approximation is exact in the limit $N_c\to\infty$. In practice,
however, it seems to be valid within a factor 2-3\cite{gimenez}.}
to estimate the value of
the dimension six condensate\cite{svz}, one indeed finds a value consistent
with that above, within a factor of two. This suggests an algorithm for
systematically improving the calculation. Adding more resonance poles to the
rational parametrization
\rf{ratva}, one can constrain the extra parameters so as to improve the
interpolation function at both ends. On the one hand 
one can ask that more terms of the
asymptotic expansion in $1/p^2$ be reproduced (in practice a better
determination of the dimension six and dimension eight condensates
than available at present is needed)
and on the other hand, in the Minkowski
region, one can improve the agreement with the experimental resonance
parameters such as $M_A$ and $F_A$.

\noindent{\bf 3.4 Resonance saturation of $<VAP>$ }

Let us now investigate the approximation of resonance saturation for the
three-point function $W_{\mu\nu}(p,q)$ \rf{w}. As explained above, this
approximation consists in constructing a rational interpolation function, the
parameters of which will be constrained from both the asymptotic
euclidian region and from the physical resonance region. In the asymptotic
region one should at least impose the constraints which are necessary 
to insure finiteness of the result for the low-energy constants $K_{11}$ and
$K_{12}$. As before, the minimal version of resonance saturation has as
resonance content the $\rho$ and the $a_1$. Whether this is sufficient in
the present situation is by no means  obvious and must 
be checked explicitly. We note first that $W_{\mu\nu}$
satisfies two Ward identities:
\be
p^\mu W_{\mu\nu}(p,q)=(-B_0 F^2_0)\left({q_\nu\over  q^2}-
{(p+q)_\nu\over (p+q)^2}\right),\quad
q^\nu W_{\mu\nu}(p,q)=(-B_0 F^2_0){(p+q)_\mu\over (p+q)^2}\ ,
\en
which imply that $W_{\mu\nu}$ must have  the following  
tensor structure,
\be\lbl{Wgen}
W_{\mu\nu}(p,q)={-B_0 F^2_0}\left[ 
{(p_\mu+2q_\mu)q_\nu\over  q^2l^2} -{g_{\mu\nu}\over l^2}
+ F(p^2,q^2,l^2)\,P_{\mu\nu}
+ G(p^2,q^2,l^2)\,Q_{\mu\nu} \right]
\en
with $l^2=(p+q)^2$ and
where $P_{\mu\nu}$ and $Q_{\mu\nu}$ are the two independent tensors which
vanish under contraction with  $p_\mu$ as well as with  $q_\nu$,
\be
P_{\mu\nu}=p_\nu q_\mu-p.q\, g_{\mu\nu},\quad
Q_{\mu\nu}=p^2 q_\mu q_\nu + q^2 p_\mu p_\nu-p.q\, p_\mu q_\nu-p^2 q^2
g_{\mu\nu}\ ,
\en
$F$ and $G$ are analytic functions of the three variables
$p^2$, $q^2$ and $l^2=(p+q)^2$ with appropriate poles and cuts. In the
simple approximate representation used here, $F$ and $G$ are meromorphic
functions. 

Consider now the asymptotic euclidian region. The OPE provides an
expansion  valid in the regime where all three momenta squared $p^2$, $q^2$ and
$l^2$ are large and negative. In other terms, scaling $p\to\lambda p$ and
$q\to\lambda q$ one obtains an expansion in inverse powers of the scale
parameter $\lambda$. The leading part in this expansion is controlled by the
dimension three $q\bar q$ condensate and scales as $1/\lambda^2$ (this
scaling behaviour is exact as anomalous dimensions cancel out). We have
also worked out the subleading part in this expansion, 
scaling as $1/\lambda^4$, 
which is proportional to the so-called mixed condensate,
\bea\lbl{wasy}
&&\lim_{\lambda\to\infty} W_{\mu\nu}(\lambda p,\lambda q)=
{<\bar u u>\over \lambda^2}\left\{
{(p_\mu+2q_\mu)q_\nu\over  q^2 l^2}-{g_{\mu\nu}\over l^2}
+{p^2-q^2-l^2\over  2p^2q^2l^2} P_{\mu\nu}
-{1\over  p^2q^2l^2}\, Q_{\mu\nu}\right\}\\ \nonumber
&&-{<\bar u\sigma_{\mu\nu}G^{\mu\nu} u>\over 6\lambda^4}\left\{
\left( {l^2\over  p^4q^4}+{p^2+q^2\over  p^4l^4}
-{p^2+q^2\over  q^4l^4}\right)\,P_{\mu\nu}-
2\left( {1\over  p^4l^4}-{1\over  q^4l^4}\right)\,Q_{\mu\nu}\right\}
+O\left({1\over \lambda^6}\right)\ .
\ena
There are also $O(\alpha_s)$ corrections which we have not evaluated. 
With the minimal resonance content, it proves  perfectly possible to match
the leading part of this asymptotic expansion. 
This may be viewed as a surprise as this matching is much more
constraining than that for the two-point function $<VV-AA>$. We
have now three independent variables instead of just one and two independent
amplitudes $F$ and $G$. In fact, this asymptotic constraint entirely
determines the rational approximant up to two constants, $a$ and
$b$
\be\lbl{GF}
F(p^2,q^2,l^2)={(p^2-q^2-l^2+2a)\over  2(p^2-\mv^2)(q^2-\ma^2)l^2 },\quad
G(p^2,q^2,l^2)={-q^2+b          \over  (p^2-\mv^2)(q^2-\ma^2) q^2 l^2}\ .
\en
Note that the effective lagrangian model used in 
refs.\cite{b+u} \cite{dhw} is inconsistent with
this asymptotic constraint. Indeed, using this model one finds
\be\lbl{GFur}
F(p^2,q^2,l^2)={1\over  F^2_0}\left({F^2_V-2F_VG_V\over(p^2-\mv^2)l^2}
                                  - {F^2_A\over        (q^2-\ma^2)l^2}
				  \right),\quad
G(p^2,q^2,l^2)={1\over  F^2_0}{-2F_VG_V\over  (p^2-\mv^2)q^2l^2}\ .
\en
Comparing with the QCD prediction in the asymptotic region, eq.\rf{wasy} one
observes that the scaling behaviour is correct but one does not match the
individual terms exactly. Worse than that, this model for 
$<VAP>$ does not satisfy the weaker constraint to reproduce the correct QED
divergence in the calculation of $2K_{11}\pm K_{12}$ (in other terms, after
taking due account of the counterterms, the result for $K_{11}$ and
$K_{12}$ is infinite in this model). The reason for these
problems is that no $\pi\rho a_1$ coupling has been introduced. Allowing for
such couplings one can recover the result of 
eq.\rf{GF}. If one now attempts to match not only the
leading asymptotic terms but also the subleading ones in the expansion
\rf{wasy} one finds that this is no longer possible unless the resonance
content is enlarged. Obviouly, for instance, in order to produce terms
proportional to $1/l^4$ we need not only the $\pi$ but also the $\pi'$
resonance, coupling to the pseudoscalar current. 

Returning to the minimal
model, it turns out that there are further chiral symmetry constraints as
well as asymptotic constraints. 
These constraints concern the pion to vacuum matrix elements that one
can deduce as residues of the pion poles in $<VAP>$. 
The minimal model will prove capable of obeying all
these constraints and the two constants $a$ and $b$ will be determined.
Consider chiral constraints first: the pion to vacuum matrix element
$<0\vert VA\vert\pi>$ must satisfy a soft pion theorem\cite{das2} 
\be
\hat W^{\mu\nu}(p,0)={i\over  F_0}\left(\Pi^{\mu\nu}_{V^3}(p) -
\Pi^{\mu\nu}_{A^3}(p)\right)\ ,
\en
Using the rational parametrization for $<VAP>$, eq.\rf{GF},  and the 
corresponding one for $<VV-AA>$, eq.\rf{ratva},  
one finds that the soft pion theorem is
obeyed provided the first Weinberg sum rule holds and the parameters $a$ and
$b$ satisfy
\be\lbl{a-b}
a-b={1\over  F^2_0}(M^2_VF^2_A-M^2_AF^2_V)=-(\mv^2+\ma^2) \ .
\en
where the second equality follows from imposing the second Weinberg sum rule. 
A second soft pion theorem associated with the pion to vacuum matrix element 
$<0\vert VP\vert\pi>$ 
is satisfied without bringing new constraints. 

We are thus left with a single arbitrary constant, $b$. 
The physical meaning of this constant  
is made clear by  identifying
the vector form-factor of the pion.
For that purpose, let us 
consider the residue of the pion pole $(p-l)^2=0$ in $\hat W_{\mu\nu}$, 
\be
\lim_{(p-l)^2\to 0} (p-l)^2 \hat W_{\mu\nu}(p,l)=
iF_0(p_\mu-2l_\mu)(p_\nu-l_\nu)\,F_V(p^2)\ .
\en
Here, the function $F_V(p^2)$ is the vector form factor of
the pion defined in a standard way 
\be
<\pi^a\vert V^b_\mu\vert\pi^c>=if^{abc}(p^a_\mu+p^c_\mu)\, F_V(p^2)\ .
\en
Using our model for $\hat W$, we obtain the following expression for 
$F_V(p^2)$
\be
F_V(p^2)=1-{b p^2\over 2\ma^2(p^2-\mv^2)}\ .
\en
It is thus  tempting to determine $b$ in order to reproduce the standard
VMD form of the vector form factor of the pion. Amusingly, this
property need not be imposed by hand here, but can be deduced
from the operator product expansion\footnote{
This actually holds only for the minimal rational approximation.}. 
Consider, indeed, the pion matrix
element of the product of the vector and pseudoscalar currents,
\be
\check W_\mu(p,l)=\int d^4x\,{\rm e}^{ipx}<0\vert T V^1_\mu(x) P^3(0)
\vert\pi(l)>\ .
\en
The OPE implies that for large euclidian values of $p^2$, 
$\check W_\mu(p,l)$ behaves as
\be
\check W_\mu(p,l)=B_0F_0\left\{ {p_\mu\over  p^2} 
+ O({1\over  p^3}) \right\}\ .
\en
This is to be compared with the result obtained from the rational
parametrization of $<VAP>$:
\be
\check W_\mu(p,l)=B_0F_0\left\{{p_\mu\over  p^2}\left[1+{p\cdot l\over 
p^2}(-2+{b\over \ma^2})\right]+{l_\mu\over  p^2}\left[2-{b\over \ma^2}\right]
+O({1\over  p^3})\right\}\ .
\en
Clearly, this will match with the OPE result provided one takes
\be\lbl{bval}
b=2\ma^2\ ,
\en
which is the same value that also insures
that the pion form factor satisfies VMD exactly. Furthermore, 
one can work out the asymptotic
expansion of the pion to vacuum matrix element $\hat W_{\mu\nu}(p,l)$ when
$p^2\to-\infty$, 
\be\lbl{Whasy}
\hat W_{\mu\nu}(p,l)=
iF_0\left\{ 
{-p_\mu l_\nu\over  p^2}+
{g_{\mu\nu}p\cdot l-p_\nu l_\mu\over  p^2}\left(
1+{p\cdot l\over  p^2}\right)
+{p^2l_\mu l_\nu-p\cdot l\, p_\mu l_\nu\over  p^4}
+O({1\over  p^3}) \right\}\ .
\en
It can be checked that this result is exactly 
reproduced in our model provided 
eq.\rf{a-b} holds. Inserting this asymptotic expansion into eq.\rf{whint}
one verifies explicitly that the UV divergence cancels out with the
counterterm contribution.

At this point, we have verified that the rational parametrization of $<VAP>$
in terms of two resonance poles (and the pion pole)
is capable of matching  the leading terms
in the  asymptotic expansion of $<VAP>$, as well as the leading asymptotic 
terms of the
two related pion to vacuum matrix elements, while obeying all the chiral
symmetry constraints. Using this parametrization 
(see \rf{Wgen}, 
\rf{GF}, \rf{a-b}, \rf{bval}) it is not difficult to compute the momentum
integrals in \rf{wint}, \rf{whint} and match the result with that of the
chiral expansion for small values of external momenta. 
One finds that 
the QED infinities do cancel out exactly, and one obtains the following 
finite results 
for the low
energy constants $K_{11}$ and $K_{12}$:
\bea\lbl{K11}
K^r_{11}(\mu)=\fac{1\over 8}\Bigg\{&&(\xi-{3\over 2})\log{\mu^2\over \mv^2}
-(\xi+3)\log{\mu^2_0\over \mv^2}-\xi-{3\over 4}\\ \nonumber
&&+3\log z\left( {(z+1)^2 \over 2(z-1)^3}+1\right)+{3z(z-3)\over (z-1)^2}
\Bigg\}
\ena
where
\be
z={\ma^2\over \mv^2}
\en
and  
\be\lbl{K12}
K^r_{12}(\mu)=\fac{1\over 4}\Bigg\{(\xi-{3\over 2})\log{\mu^2\over \mv^2}
-\xi\log{\mu^2_0\over \mv^2}-\xi+{7\over 4}
+ {3\over 2}\log z\, {z+1\over (z-1)^2} -{3z\over  z-1}\Bigg\}\ .
\en
The dependence upon the chiral renormalization scale $\mu$ agrees with that
derived in ref.\cite{urech} (see also \cite{neufeld}) 
in the gauge $\xi=1$. As was anticipated, one
observes that $K_{11}$ and $K_{12}$ also depends on the QCD renormalization 
scale $\mu_0$. The result on this dependence agrees with ref.\cite{bij2}
concerning $K_{11}$ but not $K_{12}$.
We stress that once the assumption of saturation from a
single multiplet of vector and of axial-vector resonances is made, the
result for $K_{11}$ and $K_{12}$ is uniquely determined, in much the same
way as it was for the $O(p^2)$ constant $C$, in terms of just $M_A$ and
$M_V$. We also note that once the
expression for $K_{12}$ is obtained, the three-point function $<VAP>$ may be
used for getting $K_{13}$ via  \rf{2k12+k13}. We checked that the
same result for $K_{13}$ 
which was found before from a two-point function (i.e. eq.\rf{K13},
plus resonance saturation) was indeed recovered. 

\noindent{\bf 3.5 Phenonenological implications}

Our construction of the Green's function $<VAP>$ has phenomenological
implications which are interesting to consider
in order to assess the reliability of the result for $K_{11}$ and $K_{12}$. 
Let us firstly consider the
predictions 
in the region of low momenta and then, further away, in
the resonance region. 
For small momenta, comparing with the chiral
expansion of $<VAP>$ (which can be found in ref.\cite{gl84}), 
one finds that the model reproduces the vector meson
dominance formulas for the constants $l^r_5(\mv)$ and $\l^r_6(\mv)$
which are known to be reasonably accurate \cite{gl84}. 
Obviously, however, the model
does not generate the logarithmic singularity caused by the pion loop.
It is perfectly feasible to improve on this 
by taking a more sophisticated form for
the vector meson propagator. However, this will  complicate the
computation of the photon loop integral while bringing corrections which
are subleading in the large $N_c$ counting, and should thus be rather small.  
Concerning the resonance sector, for the $\rho$
meson,  the model is found to embody 
a reasonable value for the $\pi\pi$ decay width.
For the $a_1$ resonance, 
the model predicts a value for the $\rho\pi$ decay width of the order
of $200$ MeV, which is within 50\% of the experimental value. 
A surprising result emerges for the 
radiative width, $a_1\to\gamma\pi$. Extracting the decay amplitude 
from $<VAP>$ using the reduction formula, 
one obtains the following expression 
\be
{\cal T}(a_1(q)\to\gamma(p)\pi)=(\epsilon.p\,\epsilon'.q-p.q\,
\epsilon.\epsilon')
{F_0\over  M_A F_A}\,{2M^2_V- M^2_A\over  2M^2_V}\ .
\en
Because of the factor $2M^2_V- M^2_A$, this amplitude is strongly suppressed
and one obtains
a value of a few tens of KeV for the width, 
which is one order of magnitude smaller than the
experimental value obtained in ref.\cite{zielinski}. Clearly, since the
$a_1$ pole is rather far from the integration region relevant for the
evaluation of the LEC's $K_{11}$ and $K_{12}$ it is plausible that these
constants could be reasonably well evaluated using the rational approximant
for $<VAP>$ and at the same time, the properties of the $a_1$ need not be
too precisely reproduced by the approximant. Still, a mismatch by one order
of magnitude would be disturbing.  
A closer look at the
literature, however,  reveals that a small value for the radiative width is 
not ruled out. Indeed,
a recent photoproduction experiment has found no trace of $a_1$ production 
\cite{condo93} ( while observing very clean 
evidence for $a_2(1320)$ production). 
In this experiment, the photon is on shell, which is not
necessarily exactly satisfied in the experiment of ref. \cite{zielinski}
which is a Primakov-type experiment. Our amplitude is very strongly energy
dependent and it is nearly vanishing only for exactly massless photons. 
At any rate, if the $a_1$ radiative width turns out to be unsuppressed, this
would be clear evidence for the necessity of including further resonances in
the parametrization of $<VAP>$ and would invalidate the above 
estimate of $K_{11}$ and $K_{12}$. On the contrary, confirmation of the
suppression would be a nice experimental support for these estimates.

\noindent{\bf 3.6 Some remarks on $K_1$ to $K_6$}

Let us discuss here some aspects of the convolution representations for the
constants $K_1$ to $K_6$, with some emphasis on the combination $K_5+K_6$ 
which is of interest for DT violation. 
In order to illustrate the procedure, consider the
correlator
\bea\lbl{lrqq}
\Pi^{abcd}_{\alpha\beta}(p)&&=
\int d^4x\, d^4y\, d^4z\, e^{ip{x+y\over2}}
<0\vert T\left(A^a_\alpha(x) A^b_\beta(y)-
V^a_\alpha(x) V^b_\beta(y)\right)Q^c_V(z) Q^d_V(0)\vert 0>\\ \nonumber
&&=g_{\alpha\beta}\,\Pi^{abcd}_0(p^2)+p_\alpha p_\beta\, \Pi^{abcd}_1(p^2)\ .
\ena
We will concentrate in the following on the first tensor
component, $\Pi^{abcd}_0$. 
Using the chiral lagrangian, we can compute its chiral expansion,
which involves a photon loop as well as a pion loop. Choosing flavour
indices in order to isolate $K_5+K_6$, one obtains,
\bea\lbl{pi0}
&&\Pi^{1425}_0(p^2)=iF^2_0\,\Bigg\{K^r_5(\mu)+K^r_6(\mu)
-{3\over2}K^r_{13}(\mu)+
3 K^r_{12}(\mu)\\ \nonumber
&&+{1\over16\pi^2}\,{3\over4}
\left(\log{\mg^2\over\mu^2}+{\xi-1\over4}\left[
\log{\mg^2\over\mu^2}+{\xi\log\xi\over\xi-1}-{1\over2}\right]
+{C\over F^4_0}\left[-\log{-p^2\over\mu^2}+1\right]\right)\Bigg\}+O(p^2)\ .
\ena
This expression allows one to deduce the scale dependence of $K_5+K_6$,
utilizing the previous results for $K_{12}$ (eq.\rf{K12}) and $K_{13}$
(eq.\rf{K13}), 
\be
\mu{d\over
d\mu}(K^r_5(\mu)+K^r_6(\mu))={1\over16\pi^2}\left({9\over4}-{3C\over2F^4_0}
\right) ,
\en
which reproduces the result of ref.\cite{urech} with the new feature that it
is found not to depend on the gauge parameter $\xi$ 
(we note in passing that that the correlator $<V_\alpha V_\beta Q_V Q_V >$ is
finite at one loop, which immediately implies that $d/d\mu(
K^r_{13}(\mu)+2K^r_{14}(\mu))=0$ ). 
In a similar way, it is easy to pick up two sets of flavour indices such
that the chiral expansion of the correlation function will involve the
combinations $K_1+K_2$, 
\be
\Pi^{3388}_0(p^2)=4iF^2_0\left\{K^r_1(\mu)+K^r_2(\mu)+{1\over3}(K^r_5(\mu)+
K^r_6(\mu))+{3C\over4F^4_0}\left[-\log{-p^2\over\mu^2}+1\right]\right\}
\en
and $-2K_3+K_4$,
\be
\Pi^{3838}_0(p^2)=2iF^2_0\left\{-2K^r_3(\mu)+K^r_4(\mu)+{2\over3}(K^r_5(\mu)+
K^r_6(\mu))+{3C\over2F^4_0}\left[-\log{-p^2\over\mu^2}+1\right]\right\}\ .
\en

The next step is to
express the correlator above \rf{lrqq} 
by taking functional derivatives of the QED and
QCD actions. One observes, firstly, that the counterterms make a
contribution which has the following form,
\be
\Pi^{abcd}_{\alpha\beta}(p)\vert_{counter}=
iZ_2(f^{acl}f^{dbl}+f^{adl}f^{cbl})\left(\Pi^{\alpha\beta}_{A^3}(p/2)-
\Pi^{\alpha\beta}_{V^3}(p/2)\right)\ .
\en
From this, one deduces the finite convolution representation
\bea\lbl{conv4}
\Pi^{1425}_{\alpha\beta}(p)&&=-i\int {d^4k\over(2\pi)^4}
D_{\mu\nu}(k)\left[\Pi^{\alpha\beta\mu\nu}_{A^1 A^4 V^2 V^5}(p,k)-
\Pi^{\alpha\beta\mu\nu}_{V^1 V^4 V^2 V^5}(p,k)\right]\\
\nonumber
&&-iF^2_0{3\over4}Z_2\left[g_{\alpha\beta}-{p_\alpha p_\beta\over
p^2}\right](1+O(p^2))\ ,
\ena
in terms of the QCD 4-point functions,
\be
\Pi^{\alpha\beta\mu\nu}_{A^a A^b V^c V^d}(p,k)=
\int d^4x d^4y d^4z\, 
{\rm e}^{ip{x+y\over2}+ikz}<0\vert T
A^a_\alpha(x) A^b_\beta(y))V^c_\mu(z) V^d_\nu(0)\vert 0>
\en
and the similar correlator with $A^a$, $A^b$ replaced by $V^a$, $V^b$.

The analogous representations for $\Pi^{3388}_{\alpha\beta}$ and
$\Pi^{3838}_{\alpha\beta}$ have no contribution from the counterterms
implying that the convolution integrals must be finite. 
One must extract the tensor component $g^{\alpha\beta}$ from the 
four dimensional integral and 
one obtains $K_5+K_6$ using eq.\rf{pi0}.   

The part proportional to $\xi-1$ in the convolution integral above  \rf{conv4}
can be worked out in a model independent way. 
Indeed, the contraction of $k^\mu k^\nu$ with the 4-point function
can be expressed in terms of the 2-point function $<VV-AA>$
using the chiral Ward identities. In this manner, the dependence upon $\xi$ of 
$\Pi^{1425}_0(0)$ can be  exhibited as,
\be
\Pi^{1425}_0(0)=\Pi^{1425}_0(0)\vert_{\xi=0}+
{iF^2_0\over16\pi^2}{3\over16}\,\xi\,\left\{
\log\xi+\log{M^2_\gamma\over\mu^2_0}-4
 +{3\over\pi F^2_0}\intif
dx\,\log{x\over\mu^2_0}\left[\rho_V(x)-\bar\rho_A(x)\right]\right\}
\en
in which the expression of the counterterm constant $Z_2$ (see \rf{ZZb}) 
has been used. 
This formula, together with the expressions of $K_{12}$ and $K_{13}$
\rf{K12}, \rf{K13} allows one to show 
that the combination $K_5+K_6$ is independent of the
gauge parameter $\xi$. 
Furthermore, the dependence on the QCD renormalization scale $\mu_0$ is
found to be entirely absorbed by that of the parameter $K_{12}$ such that
$K_5+K_6$ is independent of $\mu_0$. 
In a similar way, one can also show that the two
combinations $K_1+K_2$ and $-2K_3+K_4$ are gauge independent and $\mu_0$ 
independent.
Correlation representations for all of $K_1$ to $K_6$ can be obtained by
further considering the correlators $<A^a_\alpha A^b_\beta Q^c_L Q^d_R>$
from which one can isolate $K_2$, $K_4$ and $K_6$. The derivation exactly
parallels the discussion of the correlator \rf{lrqq} and will not be
detailed. Let us simply mention that there will be no contribution from the
counterterms, implying that the convolution integrals must be finite. Again,
the dependence on the gauge parameter $\xi$ can be worked out explicitly and
it turns out that the six parameters $K_1$ to $K_6$ are independent of the
value of $\xi$. 

In order now to obtain phenomenological estimates, for instance for
$K_5+K_6$, we must construct a resonance saturated model for the QCD 4-point
function $<(AA-VV)VV>$. 
The asymptotic constraint to be satisfied in this case is that the
convolution integral must be finite in the particular gauge $\xi=0$ (Landau
gauge). 
The model must also obey a set of chiral Ward identities. The
simplest way to implement these in the present situation is to first write
down a chirally invariant effective lagrangian for the resonances. 
For this purpose, one may
use, for instance, the convenient method advocated in ref.\cite{egpr}. However,
it will be necessary to 
investigate more couplings than were considered in this
reference (where only those couplings actually contributing to the LEC's
$L_i$ were discussed). We already pointed out that in order to fulfill the 
asymptotic constraints associated with $<VAP>$, vertices associated with
$a_1\rho\pi$ couplings were needed. Here, we can also expect contributions
associated with $\omega\rho\pi$ coupling (its contribution 
to $\Delta M^2_K$ was considerered
long ago \cite{socolow}) but possibly also from $b_1\omega\pi$ coupling. Also
four-particle couplings and exchanges of scalar or tensor mesons could be
considered. It is clear that a rather extensive study is necessary in order
to carefully investigate all these possibilities. This study 
will not be undertaken here. We have simply 
performed the exercise to reevaluate the LEC $K_{13}$, which can
be isolated from the 4-point function $<V^a_\alpha V^b_\beta Q^c_L Q^d_R>$,
using the resonance effective lagrangian of Ecker et al.\cite{egpr}. The
number of graphs to consider is strongly reduced by working in the Landau
gauge (the gauge dependence can be worked out explicitly as above).
The calculation is much more tedious than the one based on a two-point
function in sec.3.1, but one indeed recovers exactly the same result.

\noindent{\large\bf 4. Flavour symmetry breaking and the constants
$K_7,...,K_{10}$ }

\noindent{\bf 4.1 The constants $K_7$ and $K_8$}

We can isolate the contributions of the two constants $K_7$ and $K_8$ 
by taking the derivative with respect to $m_s$ of the Q-correlators
$<Q_{V^3} Q_{V^3}>$ and $<Q_{A^3} Q_{A^3}>$ already considered. After
a simple calculation, one obtains from the chiral lagrangian,
\be
{d\over  d\ms}\Pi^Q_{V^3}(0)=2B_0 F_0^2\left\{
2(K^r_7(\mu)+K^r_8(\mu))-\fac{C\over F_0^4}\Big(1
+\log{\mk^2\over \mu^2}\Big) \right\}+ O(m_q)
\en
and
\be
{d\over  d\ms}\Pi^Q_{A^3}(0)=2B_0 F_0^2\left\{
2(K^r_7(\mu)-K^r_8(\mu))+\fac{C\over F_0^4}\Big(1
+\log{\mk^2\over \mu^2}\Big) \right\} +O(m_q)\ .
\en
These expressions display the scale dependence of  $K^r_7$ and
$K^r_8$. In particular $K_7$ is finite and scale independent. One may then
use the convolution representation for the Q-correlators $\Pi^Q_{V^3}$ 
and $\Pi^Q_{A^3}$ in terms of QCD correlators but this does not lead
to an expression which can easily be evaluated. 
However, it is not difficult to show that the derivatives with
repect to the strange quark mass of the isospin one vector and axial-vector
spectral functions are suppressed in the large $N_c$ counting. This is
because the graphs which contribute involve one extra (strange) quark loop
with respect to the leading order graphs. The fact that $K_7$ and $K_8$
are subleading in $N_c$ could, of course, have been suspected from the
double trace structure of the corresponding lagrangian terms (see
ref.\cite{bij2} for more details on this aspect). One may assume, 
then, that $K_7$ and $K^r_8$ are suppressed for scales $\mu\simeq\mv$ 
and infer the following estimate,
\be
K_7\simeq0,\qquad K^r_8(\mu)\simeq\fac{C\over 2F_0^4}
\,\log{\mv^2\over \mu^2}\ .
\en

\noindent{\bf 4.2 Constraints on symmetry breaking spectral function
differences}

In order to discuss the constants $K_9$ and $K_{10}$, we consider the
correlator
\be\lbl{dvq}
\Delta\Pi^Q_V(p^2)={-3i\over 2(\mhat-\ms)<\bar u u>}\int dx {\rm e}^{ipx}
<0\vert T(Q^3_V(x) Q^3_V(0) - Q^8_V (x)Q^8_V(0))\vert0>_{lin}\ .
\en
and a similar correlator $\Delta\Pi^Q_A$ with $V$ replaced everywhere
by $A$ in the above formula. 
The normalization factor, in front of the integral is introduced for
convenience.
The subscript $lin$ means that we must 
keep only terms linear (or logarithmic) in the quark masses and drop
the quadratic or higher order terms (the reason for this will be given below). 
Using the definition of the Q-currents in terms of the QCD vector 
or axial currents and the spectral representation of the correlator
$\Pi^{\mu\nu}_{V^3}(q)$ (see \rf{specv}) and the analogous definition for
$\Pi^{\mu\nu}_{V^8}(q)$, 
the following spectral function difference will appear, 
\be\lbl{deltaro}
\Delta\rho_V(x)={-3\over 2(\mhat-\ms)<\bar u u>}\Big(\rho_{V^3}(x)
-\rho_{V^8}(x)\Big)_{lin}\ .
\en
Again, we can also define $\Delta\rho_A(x)$ by replacing
$\rho_{V^a}$ by $\rho_{A^a}$ everywhere in \rf{deltaro}. 
In the latter case, as we are away from the chiral limit,
there is a second spectral function $\sigma_A$ 
(see \rf{speca} ) to be considered and we define
\be
\Delta\sigma_A(x)={-3\over 2(\mhat-\ms)<\bar u u>}\Big(\sigma_{A^3}(x)-
\sigma_{A^8}(x)\Big)_{lin}\ .
\en
With these definitions
$\Delta\rho_V$ and $\Delta\rho_A$ must satisfy sum rules analogous to the 
two Weinberg sum rules. Indeed, at large $-q^2$ , the leading term 
in the asymptotic behaviour of  
the QCD correlation function difference $<V^3V^3-V^8V^8>$ 
is given by\cite{svz} 
\be
\lim_{p^2\to-\infty}
p^4\left(\Pi^{\mu\nu}_{V^3}(p^2) -\Pi^{\mu\nu}_{V^8}(p^2)\right)_{lin}=
{2\over 3}(\mhat-\ms)<\bar u u>\ \left(p^\mu p^\nu-g^{\mu\nu}p^2\right)
.
\en
This implies the two sum rules,
\be\lbl{srvec}
\intif dx\,\Delta\rho_V(x)=0,\qquad
I_V\equiv{1\over \pi}\intif dx\,x\,\Delta\rho_V(x)=1\ .
\en
For these sum rules to hold, it is essential to drop the quadratic mass
terms in the definition of $\Delta\rho_V(x)$. Otherwise, the first sum rule
would still be valid but the second one would diverge.
The first of the above sum rules was established
long ago\cite{das2} (and is customarily referred to as the DMO sum rule) 
while, curiously,  we could not find a trace of the
second one in the literature. Analogous sum rules also hold for the
axial current spectral functions
\be\lbl{srax}
\intif dx\,\Delta\rho_A(x)=0,\qquad
I_A\equiv{1\over \pi}\intif dx\,x\,\Delta\rho_A(x)=1\ ,
\en
together with
\be\lbl{srax1}
{1\over \pi}\intif dx\,\Delta\sigma_A(x)=2 \ .
\en
The latter sum rule is in fact saturated by the pseudo-Goldstone boson 
contributions to $\Delta\sigma_A$, 
\be\lbl{dsig}
\Delta\sigma_A(x)=2{M^2_\pi \delta(x-M^2_\pi)-M^2_\eta
\delta(x-M^2_\eta)\over  M^2_\pi-M^2_\eta}+\Delta\bar\sigma_A(x)\ .
\en
Because the divergence of the axial current is linear in  the quark
masses, the   piece $\Delta\bar\sigma_A(x)$ is of higher order in the quark 
expansion than the first term in \rf{dsig} and must be dropped. 

Instead of the neutral vector currents $V^3$ and $V^8$ 
(or the corresponding axial vector currents) one could equally well
employ charged currents $V^{ud}$ and $V^{us}$
\be
V^{ud}_\mu=\bar u\gamma_\mu d,\quad
V^{us}_\mu=\bar u\gamma_\mu s ,
\en
(and their hermitian conjugates). 
In particular, the sum rules \rf{srvec} and \rf{srax} hold unchanged
for the spectral function difference of the charged vector current
$\Delta\rho^+_V$ defined as follows,
\be\lbl{drop}
\Delta\rho^+_V(x)={-(\rho_{V^{ud}}-\rho_{V^{us}})\over (\mhat-\ms)
<\bar uu>}
\en
and the corresponding definition for the axial-vector case. The charged
vector currents are not conserved but this effect is proportional to the
quark masses and can be ignored here.

\noindent{\bf 4.3 Sum rules for $K_9$ and $K_{10}$}

Let us now return to the Q-current correlator $\Delta\Pi^Q_V(p^2)$ 
defined in \rf{dvq}. As before, we first compute this correlator  
from the chiral lagrangian, up to chiral order 
four. Setting   $p^2 =0$, one obtains: 
\be\lbl{dvchir}
\Delta\Pi^Q_V(0)=8\left[ K^r_9(\mu)+K^r_{10}(\mu)-\fac
{3\over 4}\,Z^0(\mu)\right]+O(m_q)\ ,
\en
where
\be\lbl{z0mu}
Z^0(\mu)=
{C\over F_0^4}
{\mk^2\log(\mk^2/\mu^2)-\mpi^2\log(\mpi^2/\mu^2)\over \mk^2-\mpi^2}\ 
\en
is the pion (and kaon) tadpole contribution. 
In terms of QCD currents, we obtain the convolution representation
\be\lbl{dvconv}
\Delta\Pi^Q_V(0)
={-3\over 2(\mhat-\ms)<\bar uu>}\int{d^4k\over (2\pi)^4}\,
\left(\Pi^{\mu\nu}_{V^3}(k)-\Pi^{\mu\nu}_{V^8}(k)\right)
\left(-iD_{\mu\nu}(k)\right) +Z_s-Z_2\ .
\en
Next, inserting the spectral function representation for the correlator
$<V^3V^3-V^8V^8>$ yields
\be\lbl{dvspec}
\Delta\Pi^Q_V(0)
=\fac{3\over \pi}\intif dx\,x\, \Delta\rho_V(x)
\Big[\Gamma(\epsilon)+\log4\pi+{1\over 3}-\log{x}\Big] +Z_s -Z_2 \ .
\en
Some remarks are in order here concerning the ultraviolet divergence.
From the large momentum
expansion of the correlator $<V^3V^3-V^8V^8>$, one observes 
that there are two terms 
which will lead to a divergence in the photon loop integral eq.\rf{dvconv},
\be\lbl{dvasy}
\Pi^{\mu\nu}_{V^3}(p)-\Pi^{\mu\nu}_{V^8}(p)=
\left(p^\mu p^\nu-g^{\mu\nu}p^2\right)
\left\{ {2\over 3p^4}(\mhat-\ms)<\bar uu>
\Big[1+{\alpha_s(p^2)\over 3\pi}+...\Big]+O({1\over p^6})\right\}
\en
(retaining only terms linear in the quark masses). 
The first term in the square bracket in eq.\rf{dvasy} generates a
$1/\epsilon$ pole in dimensional regularization. This pole appears
explicitly in the spectral representation \rf{dvspec}. 
Using the 
expression of the QED counterterms $Z_s$ and $Z_2$ \rf{ZZb} 
one easily verifies that the
$1/\epsilon$ 
infinities cancel exactly. This cancellation requires that the 
second sum rule \rf{srvec} be satisfied. It is important to keep this 
point in mind in phenomenological applications based on resonance saturation. 
In particular, it is not consistent to ignore flavour symmetry breaking in the
resonance sector as has been done in ref.\cite{b+u}.
If one does not impose that this sum rule be obeyed, the result will be
infinite. 
Now the second term in the square bracket \rf{dvasy}, proportional to
$\alpha_s$, also generates a divergence in the photon loop integral, of the
form $\log\epsilon$ in dimensional regularization  
(such a smooth singularity 
is the result of renormalization-group improvement). This divergence should
be removed by minimal subtraction. 
This means that the integral
$\intif dx x \log{x}\Delta\rho_V(x)$ appearing in eq.\rf{dvspec} is, 
in fact, divergent 
(the same holds true for the axial spectral function $\Delta\rho_A$). Since
the integral $\intif dx x \Delta\rho_V(x)$ is convergent, this suggests that
the spectral function difference behaves as $\Delta\rho_V\sim1/x^2\log^2x$
asymptotically.
This remark can be used to minimally subtract the $\log\epsilon$ singularity
in the spectral representation, which should be done in practice 
if one were
to use realistic spectral functions.  
The simple models which will be considered
below do not lead to such singularities, so we will ignore this subtlety
in the following.

In a similar way, one obtains for the corresponding axial current
correlator, firstly from the chiral lagrangian
\be\lbl{dachir}
\Delta\Pi^Q_A(0)=8\Bigg[ K^r_9(\mu)-K^r_{10}(\mu)
+{\xi\over 16\pi^2}\, {M^2_\eta\log(M^2_\eta/\mu^2)-
\mpi^2\log(\mpi^2/\mu^2)\over  4(M^2_\eta-\mpi^2)}\\ \nonumber
+\fac{3\over 4}\,Z^0(\mu)\Bigg]+ O(m_q)
\en
and secondly from the QCD action
\bea\lbl{daqcd}
\Delta\Pi^Q_A(0)
=&&{-3\over 2(\mhat-\ms)<\bar uu>}\int{d^4k\over (2\pi)^4}\,
\left(\Pi^{\mu\nu}_{A^3}(k)-\Pi^{\mu\nu}_{A^8}(k)\right)
\left(-iD_{\mu\nu}(k)\right) -Z_s-Z_2\\ \nonumber
=&&\fac{3\over \pi}\intif dx\,x\, \Delta\rho_A(x)
\Big[\Gamma(\epsilon)+\log4\pi+{1\over 3}-\log{x}\Big]\\ \nonumber
&&-\fac{1\over \pi}\intif dx\,\Delta\sigma_A(x)
\Big[(\xi+3)\Big(\Gamma(\epsilon)+\log4\pi+1-\log{x}\Big)-2
\Big] -Z_s -Z_2\ .
\ena
Again here, the sum rules \rf{srax} and \rf{srax1} 
for $\Delta\rho_A$ and $\Delta\sigma_A$ ensure the correct cancellation
of infinities. One notices that the pion contribution in the photon loop 
in eq.\rf{dachir} appears also in eq.\rf{daqcd} in spectral 
representation so that this contribution cancels out in the expression
for $K_9-K_{10}$. 
One also expects that the chiral logarithms 
in the function $Z^0(\mu)$ in eqs.\rf{dvchir} and \rf{dachir} 
should cancel out with similar terms present in the spectral
functions but we will not attempt to elucidate exactly how this happens
and we will keep this contribution (which is 
subleading in the large $N_c$ counting) as it is for the moment.

It is now easy
to deduce the sum rules for $K_9$ and $K_{10}$. Defining the two
integrals
\be\lbl{ZvZa}
Z_V=\intif dx\,x\,\log{x\over \mv^2}\Delta\rho_V(x),\qquad
Z_A=\intif dx\,x\,\log{x\over \mv^2}\Delta\bar\rho_A(x)
\en
where $\Delta\bar\rho_A$ is the part of the spectral function with no pion
contribution,
\be
\Delta\rho_A(x)=2{F^2_\pi\delta(x-M^2_\pi)-F^2_\eta
\delta(x-M^2_\eta)\over  (M^2_\pi-M^2_\eta)F^2_\pi}+\Delta\bar\rho_A(x) \ .
\en
(and the worrisome asymptotic tail related to $\alpha_s$ is assumed to be
removed).
We find the following result for $K_9$:
\be\lbl{K9}
K^r_9(\mu)= \fac{1\over 8}\Bigg\{\xi\log{\mu^2\over \mu^2_0}-\xi
-{3\over 2}(Z_V + Z_A)\,\Bigg\}
\en
and for $K_{10}$,
\be\lbl{K10}
K^r_{10}(\mu)=\fac{1\over 8}\Bigg\{-\xi\log{\mu^2\over \mv^2}+
(\xi+3)\log{\mu^2_0\over \mv^2}+\xi
+{3\over 2}(Z_A-Z_V)+1+6Z^0(\mu)\Bigg\}\ .
\en
The same expression for the constants $K_9$ and $K_{10}$ holds in which the
spectral function differences related to the neutral currents $\Delta\rho_V$
and $\Delta\rho_A$ are replaced by the charged ones $\Delta\rho^+_V$ and 
$\Delta\rho^+_A$ (see eq.\rf{drop}). The only modification to eqs.\rf{K9}
and \rf{K10} is that the tadpole function $Z^0(\mu)$ must be replaced by
$Z^+(\mu)$, with
\be
Z^+(\mu)={C\over 6 F_0^4(M^2_K-M^2_\pi)}\left\{
3M^2_\eta \log{M^2_\eta\over\mu^2}+2M^2_K \log{M^2_K\over\mu^2}
-5M^2_\pi \log{M^2_\pi\over\mu^2}\right\}\ .
\en

\noindent{\bf 4.4 Difficulties of minimal resonance saturation} 

As one can see from eqs.\rf{K9}, \rf{K10} and from 
\rf{deltamk}, \rf{6termes} below,  
in order to make a definite prediction concerning the violation of Dashen's
theorem one must be able to evaluate, in a reliable way, the 
the integrals $Z_A$ and $Z_V$ defined in \rf{ZvZa}. A priori, we will
envisage to estimate these integrals in the approximation of resonance
saturation, retaining the contributions of the lowest-lying vector meson
and axial-vector meson octets. One must first 
verify that the spectral functions
$\Delta\rho_V(x)$ and $\Delta\rho_A(x)$ approximated in this way obey
the sum rules \rf{srvec} and \rf{srax} with acceptable
values of the resonance parameters. The first sum rule for $\Delta\rho_V$
and the equivalent sum rule for $\Delta\rho^+_V$ imply, respectively,
\be\lbl{sr1}
F_{\rho^0}=(3F^2_\omega+3F^2_\phi)^{1\over 2},\qquad
F_{\rho^+}=F_{K^{*+}}
\en
where the couplings of the neutral vector mesons are defined with respect
to the electromagnetic current, i.e., $<0\vert j^{em}_\mu\vert V^0>=
M_{V^0}F_{V^0}\epsilon_\mu$. Assuming exact isospin conservation allows one
to extract the matrix elements of the currents $V^3_\mu$ and $V^8_\mu$. 
Using the $e^+e^-$ branching ratios, one obtains
\be\lbl{fv}
F_{\rho^0}=152.9\pm4,\quad
F_\omega=45.9\pm1.0,\quad
F_\phi=79.1\pm1.5\quad {\rm (MeV)}\ .
\en
If one now extracts the value of $F_{\rho^0}$ using the first relation
\rf{sr1}, one obtains $F_{\rho^0}\simeq158$.  
This is in rather reasonable agreement with the
experimental value. We can also extract the experimental 
values of the charged decay
constants from $\tau$ decay data. 
Using the most recent compilation \cite{pdg96} one gets
\be
F_{\rho^+}=146\pm2,\qquad F_{K^{*+} }=151.5\pm4.7\ .
\en
The  values of $F_{\rho^+}$ 
and $F_{K^{*+}}$ turn out indeed to be nearly equal 
as is demanded by the DMO sum rule
for $\Delta\rho^+_V$ saturated by a single resonance multiplet. This is not 
a completely trivial result if one thinks that
typical coupling constants such as $F_\pi$ and $F_K$ differ by 20\%. In
conclusion, we seem to find that minimal resonance saturation is a good
approximation as far as the first sum rule for $\Delta\rho_V$ is concerned.
Let us now examine the analogous sum rule for the axial currents.
Experimentally, the only accessible data concern the charged currents. Using
the axial version of the DMO sum rule 
for $\Delta\rho^+_A$, together with resonance saturation, implies
\be\lbl{sr1a}
F^2_{K_1(1270)}+F^2_{K_1(1400)}-F^2_{a1}=F^2_\pi-F^2_K \ .
\en
In this relation $F_\pi$ and $F_K$ are, of course, rather accurately known. 
Concerning $F_{a_1}$, an estimate can be made under the
hypothesis that the decay amplitude $\tau\to\pi^-\pi^0\pi^0$ 
(the most recent PDG\cite{pdg96} value for the branching fraction is
$9.27\pm0.14\%$) is
dominated by the $a_1$ resonance and proceeds via $a_1\to\rho\pi$ 
(in other terms, we take
$\Gamma(\tau\to a^-_1\nu_\tau)=2\Gamma(\tau\to \pi^-\pi^0\pi^0\nu_\tau)$),
which gives $F_{a_1}=165\pm13$ MeV with $M_{a_1}=1230\pm40$ MeV).
The first published data for the tau decay rate into
$K_1(1270)$ and $K_1(1400)$ by the TPC/2$\gamma$
collaboration\cite{tpc} have rather large error bars. 
Much more precise results obtained by
the ALEPH collaboration were presented very recently\cite{davier}. 
Transcribed into coupling constants these results read
\be
F^2_{K_1(1270)}=123^2\pm 63^2,\qquad
F^2_{K_2(1400)}=97^2\pm  68^2\quad {\rm (MeV}^2)\ .
\en
The errors are certainly larger than in the vector case, still it 
appears plausible that the axial DMO sum sule is  reasonably well
saturated from the lowest lying resonance contributions.

Let us now turn to the second sum rule that $\Delta\rho_V$ and
$\Delta\rho_A$ must satisfy \rf{srvec}, \rf{srax}. 
Consider the vector currents first. Computing the integral $I_V$ using minimal
resonance saturation and enforcing the first sum rule gives
\be\lbl{iv}
I_V={F^2_V(M^2_\rho-M^2_\phi)\over \F^2(M^2_\pi-M^2_K)}\ .
\en
The second sum rule states that $I_V=1$.
Unfortunately, this relation  completely fails to be satisfied as, using 
experimental values, one finds $I_V\simeq5.4$. 
Neither experimental uncertainties nor the use of the narrow width
approximation can explain such a large discrepancy. 
One source of uncertainty stems from the requirement of expanding the
numerator to linear order in the quark masses. We have assumed that the term
linear in the quark masses dominates the chiral expansion of the vector
meson masses, as is suggested by 
the success of the Gell-Mann-Okubo mass formula for the vector nonet. 
The only possible
explanation, then, is that the integral $I_V$ picks up 
significant contributions from the
energy region above 1 GeV, implying that the simplest resonance approximation
method appears to  fail in this case. 
As has been discussed in sec.3 this approximation consists in performing
a rational interpolation of the correlator $<V^3V^3-V^8V^8>$ with two poles.
In the present case, the interpolation is not capable of correctly matching
both the region of large $p^2$ and the region of small $p^2$. In order to
improve it, we must increase the number of poles. 
If one nevertheless insists on employing a single resonance multiplet, then
one must use an {\it unphysical} value for the product
$F^2_V(M^2_\rho-M^2_\phi)$  such that $I_V$ in \rf{iv} satisfies 
$I_V=1$ ( recall that 
if the sum rule  were not satisfied the result for $K_9+K_{10}$ would be
infinite). Doing this we find the following value for the integral $Z_V$
\rf{zv} which occurs in the expressions of $K_9$ and $K_{10}$: 
\be
Z^{min}_V=I_V=1\ .
\en
Equality of $Z^{min}_V$ and $I_V$ results from using 
the narrow width approximation, together with the first sum rule, and expanding
linearly in the quark masses. 
This result, however, is not stable against inclusion of higher mass
resonances.
Indeed, let us include one additional multiplet of vector resonances.
One can fit
the value of $F^2_{V'}(M^2_{\rho'}-M^2_{\phi'})$ 
as well as the splitting $F^2_{\rho'}-F^2_{\phi'}$
in order that the first two sum rules be obeyed\footnote{This is 
possible provided $F^2_\rho-3F^2_\omega-3F^2_\phi$ is negative which,
fortunately, happens to be the case.} while keeping physical values for $F_V$
and $M^2_\rho-M^2_\phi$.  This certainly improves the validity of the
interpolation function for small values of the $p^2$. 
Whether the values of the new couplings $F_{\rho'}$, $F_{\phi'}$ are
at all realistic cannot be decided since 
not much is experimentally known about these quantities.  
This improved model yields the following
evaluation of $Z_V$: 
\eject
\bea\lbl{zv}
Z^{impr}_V=1+{3(F^2_\rho-3F^2_\omega-3F^2_\phi)\mv^2\over 2(\mpi^2-\mk^2)\F^2}
&&\left({\mvp^2\over \mv^2}-\log{\mvp^2\over \mv^2}-1\right)\\ \nonumber
&&+\log{\mvp^2\over \mv^2}\left(1-{(M^2_\rho-M^2_\phi)F^2_V\over 
(\mpi^2-\mk^2)\F^2}\right)\simeq-3.9\ ,
\ena
taking $\mvp=1.6$ GeV, which differs considerably from the estimate based on
one resonance multiplet.

Turning to the axial-vector  sector now, let us evaluate the integral $I_A$
in \rf{srax}
under the approximation of minimal resonance saturation. 
Assuming the first sum rule to be satisfied, one obtains,
\be
I_A={1\over (\mpi^2-\mk^2)\F^2}\left[F^2_A(M^2_{a_1(1260)}-
M^2_{f_1(1510)})+2M^2_A(F^2_K-F^2_\pi)\right]+2,
\en
and $I_A$ should be equal to one according to the sum rule \rf{srax}.
Using experimental numbers, one finds instead $I_A\simeq 6\pm2$.
If one uses for $M_A$ and $F_A$ the values from the Weinberg sum rules
(i.e. $F_A=122$ MeV, $M_A=966$ MeV with our choice of $F_V$ and $M_V$ and
ignoring the difference between $F_0$ and $F_\pi$)
rather than the experimental ones, then one obtains $I_A\simeq 4$.
The approximation of minimal saturation is again 
found to be in conflict with the
second sum rule, if one takes physically reasonable 
values for the resonance parameters. 

At this point, the conclusion would be that it is not possible to obtain a
reliable estimate of both $Z_V$ and $Z_A$ (and consequently of the two LEC's
$K_9$ and $K_{10}$) at present, because the spectral integrals are too
slowly converging. However, if one is only interested in the amount of
Dashen's theorem violation one really needs  
to be able to estimate $K_{10}$, 
because $K_9$ happens to be multiplied by $M^2_\pi$ and will contribute
very little. Now the convolution representation for $K_{10}$ involves the
double difference $\Pi_{V^3}-\Pi_{A^3}-\Pi_{V^8}+\Pi_{A^8}$.   
The corresponding spectral functions must satisfy two sum rules obtained by
combining \rf{srvec} and \rf{srax}
\be\lbl{srk10}
\intif dx(\Delta\rho_V(x)-\Delta\rho_A(x))=0,\quad
\intif dx\,x(\Delta\rho_V(x)-\Delta\rho_A(x))=0\ .
\en
The second sum rule implies that convergence of the spectral representation
should be  faster than it is for $\Pi_{V^3}-\Pi_{V^8}$ and 
$\Pi_{A^3}-\Pi_{A^8}$ individually. In fact, one can observe that the second
sum rule in \rf{srk10}, which can be stated 
as $I_A=I_V$, is approximately satisfied in the minimal
resonance approximation as one can check from the values of $I_A$ and $I_V$
obtained above.
Experience with the Weinberg sum rules and the corresponding DGMLY sum rule
for $C$ suggests that minimal resonance saturation is likely to be
a reasonable approximation as far as $K_{10}$ is concerned.  
In this approximation, the value of
the difference $Z_A-Z_V$ to be used in eq.\rf{K10} for $K_{10}$ is given by
\be\lbl{ZamZv}
Z_A-Z_V={1\over\F^2(\mpi^2-\mk^2)}\left[
2\ma^2(F^2_\pi-F^2_K)+\fv^2(M^2_\rho-M^2_\phi)\log{\ma^2\over\mv^2}
\right]-2\left(1+\log{\ma^2\over\mv^2}\right)\ ,
\en
where the sum rule $I_A-I_V=0$ has been used to simplify the expression.

\newpage
\noindent{\large\bf 5. Consequences for Dashen's theorem violation}

Up to corrections which are quadratic in the light quark masses or quartic
in the electric charge, the contribution to the $\kp-\kz$ mass difference
proportional to $e^2$ can be written 
as a sum of six terms\cite{urech}, as follows:
\be\lbl{deltamk}
\Delta M^2_K\equiv\mkp^2-\mkz^2\vert_{em}=t_0+t_1+t_2+t_3+t_4+t_5
\en
with
\bea\lbl{6termes}
&&t_0=e^2{2C\over  F^2_0}\\ \nonumber
&&t_1={-e^2\over 16\pi^2}\left[3\mk^2\log{\mk^2\over \mu^2}-4\mk^2+
{2C\over F_0^4}\left(\mpi^2\log{\mpi^2\over \mu^2}
+2\mk^2\log{\mk^2\over \mu^2}\right)\right]\\ \nonumber
&&t_2=-e^2{16C\over F_0^4}\left[(\mpi^2+2\mk^2)L^r_4+\mk^2 L^r_5\right]\\
\nonumber
&&t_3=e^2\mpi^2\left[4K^r_8+{4\over 3}(K^r_9+K^r_{10})\right]\\ \nonumber
&&t_4=e^2\mk^2\left[ 8K^r_8+8(K^r_{10}+K^r_{11})\right]\\ \nonumber
&&t_5=-e^2\mk^2{4\over 3}(K^r_5+K^r_6)\ .
\ena
One notices first, using expressions \rf{K9}, \rf{K10} for $K_9$ and
$K_{10}$ and \rf{K11} for $K_{11}$ and the discussion in sec. 3.6
on $K_5+K_6$ that the dependence on the QED gauge
parameter $\xi$ drops out, as it should. 
In fact, the sum $t_0+t_1+t_2+t_3+t_4$ and $t_5$ are separately gauge
independent. 
The dependence upon the QCD
renormalization scale $\mu_0$ drops out in the term $t_4$ but not in the
term $t_3$.  
This indicates that the electromagnetic contribution to the $\kp-\kz$ mass
difference is not, strictly speaking, an observable quantity: only the full
mass difference is independent of $\mu_0$. 
Fortunately, this term $t_3$ makes a negligibly small contribution to the
mass difference, being proportional to $M^2_\pi$. In order to estimate
numerically the value of $\Delta M^2_K$ one needs, essentially, an estimate
of the three constants $K_{10}$, $K_{11}$ and  $K_5+K_6$ ignoring the terms
which are proportional to $M^2_\pi$ or suppressed by the Zweig rule. 
The estimate of $K_{10}$ based on resonance saturation is contained in 
the expressions \rf{K10} and \rf{ZamZv} and a similar estimate for $K_{11}$
is given in \rf{K11}.
We were not able to
evaluate $K_5+K_6$ but one notices that this combination appears 
in the expression for
$\Delta M^2_K$ with a much smaller numerical coefficient than $K_{10}$ or
$K_{11}$. Assuming that $K_5$ and $K_6$ have the order of magnitude typical
of $O(p^4)$ LEC's the lack of precise evaluation generates a rather small
uncertainty. 
Based on these estimates one obtains the following decomposition of $\Delta
M^2_K$:
\be\lbl{dash1r}
M^2_\kp-M^2_\kz\vert_{em}=(M^2_\pip-M^2_\piz)\left\{
1.02+1.13-0.20+[-2,2]10^{-2}+0.80+[-0.3,0.3]\right\}\ .
\en
where the six entries correspond to the six contributions $t_0,..,t_5$
displayed in eq.\rf{6termes}. 
Individual contributions are scale dependent: they are shown
at the CHPT scale $\mu=\mv$ and the
QCD scale $\mu_0=1$ GeV. The values of $M_A$ and $F_A$ wherever they appear
are taken from the Weinberg sum rules. 
The contribution from $t_3$ (fourth entry in eq.\rf{dash1r}) has been
estimated to lie in a range, assuming that $Z_V$ could take values between
$\pm 4$ (see sec.4.4). The last entry corresponds to the contribution of $K_5$
and $K_6$ assuming that each of these LEC's lies in the range $\pm1/16\pi^2$. 

It is interesting to verify the stability of the result if one drops
systematically all terms which are known to be subleading in the large $N_c$
counting. 
This is easily accomplished in eq.\rf{deltamk}
knowing that $F_0^2$, $L^r_5$ and $C$ are $O(N_c)$, 
$L^r_4$ is $O(1)$, 
$ K^r_5,\ K^r_6,\ 
K^r_9,\ K^r_{10},\ K^r_{11}$ are also $O(1)$ 
while $K^r_8$ is $O(1/N_c)$. The variation of $L^r_5(\mu)$ 
with respect to the scale $\mu$ is subleading in $N_c$ but, 
in practice, it is not negligible since $L^r_5$ varies by a factor of two as
$\mu$ is varied between $M_\eta$ and $M_\rho$. This raises the question of
which scale should one choose for performing a leading large $N_c$
calculation. We bypassed this difficulty by expressing directly $L_5$ in
terms of experimental data (i.e. $F_K$ and $F_\pi$) dropping the contributions
which are subleading in $N_c$ in the chiral expansions of $F_K$ 
and $F_\pi$. In this way, one obtains the scale
independent estimate, 
\be
L_5={F^2_K-F^2_\pi\over8(M^2_K-M^2_\pi)}\simeq 2.3 10^{-3}
\en
Using this result, and dropping all subleading terms
in $N_c$ we obtain the $\kp-\kz$ mass difference in the following form,
\be\lbl{dash1rNc}
M^2_\kp-M^2_\kz\vert_{em}=(M^2_\pip-M^2_\piz)\left\{
1.02+0.75-0.51+[-1,2]10^{-2}+1.20+[-0.3,0.3]\right\}\ .
\en
Interestingly, though individual terms are substantially modified 
when removing these contributions  which are 
subleading in $N_c$, the overall result is reasonably stable.
The contribution of $K_5+K_6$ still remains to be explored in detail, before
a quantitative estimate of $\Delta M^2_K$ can be provided by this approach.
Still, the results above are strongly suggestive of a rather large amount of
violation of Dashen's theorem, similar in magnitude to the results of
\cite{dhw}, \cite{perez}, \cite{bij1} or to the calculation performed on the
lattice \cite{lattice} in the quenched approximation. 

The cause for the different result obtained in ref.\cite{b+u}
(negligible DT violation) can be better understood with the help of a recent
preprint by the same authors\cite{b+u2} in which they have computed, from
exactly the same resonance lagrangian, the contributions to each of the
constants $K_i$. Firstly, their neglect of flavour symmetry breaking in the
resonance sector amounts to setting $K^r_{10}(M_V)=0$. Doing this bring the
contribution of $t_4$ to $\Delta M^2_K$ 
down from approximately one to 0.3, using our
value for $K_{11}$. However, their estimate 
\footnote{The correspondance of their results with our formulas is as
follows: one must drop the counterterm contributions and define the
convolution integrals using the chiral $\overline{MS}$ prescription. The QCD
correlators must, of course, be computed using their resonance lagrangian. 
In this way one immediately recovers that $K^r_{10}=0$ and we also checked
their results for $K_{11}$ and $K_{12}$. We could not, however, reproduce
their result  for $K_{13}$.}
of  $K_{11}$
is much smaller than ours such that $t_4$ is practically zero in 
their evaluation. As mentioned in ref.\cite{b+u2}, a remainder
contribution is needed in order to restore the correct chiral scale and QCD
scale dependences. Comparison with our results shows that this remainder is
not negligible, and may even be dominant as in the case of $K_{11}$. 

We note that neither \cite{dhw}, 
\cite{perez} or \cite{b+u} have included enough terms 
(such as $\pi\rho a_1$ coupling, see sec 3.4) in their resonance 
lagrangian such as to satisfy the asymptotic conditions needed to evaluate
{\sl separately} $K_{10}$, $K_{11}$ and $K_5+K_6$. 
A less constraining set of two sum rules were implemented in
ref.\cite{perez} which are derived by
requiring UV finiteness of $\Delta M^2_K$ (i.e. no contribution from QED
counterterms), which holds in the limit
$m_u=m_d=0$. This is obviously a minimal requirement for any realistic model
but it is not satisfied by the estimate of ref.\cite{b+u}. In
ref.\cite{perez} these two constraints are enforced by adjusting the
two values of the axial-vector coupling constant $F_{K_1}$ (assuming
$F_{K_2}=F_{K_1}$) and $F_{K'_1}$, the coupling constant of an
excited axial resonance which they find necessary to include. These two sum
rules are reminiscent of those obtained above, eq.\rf{srk10}, 
in connection with the parameter $K_{10}$,   
the difference is that they pick up additional
contributions associated with 
the combination $8K_{11}-4/3 (K_5+K_6)$ which occurs in the
expression of $\Delta M^2_K$. We recall that the two sum rules 
\rf{srk10} did not seem to require significant contributions from excited
resonances. A better experimental determination of the coupling constants
$F_{K_1}$ and $F_{K_2}$ would help to clarify these issues. 
We note also
that the calculation of ref.\cite{perez} includes contributions which are
quadratic in $m_s$, it would be of interest to know the relative importance
of these contributions.
Our result for $K_{10}$ is in reasonable agreement with
that of ref.\cite{bij2}: dropping the tadpole contribution, setting
$\mu=m_\rho$, $\mu_0=0.7$ GeV and $\xi=1$, we obtain
$K^r_{10}(\mu)=5.2\,10^{-3}$, to be compared with the value
$K^r_{10}(\mu)=(4\pm1.5) 10^{-3}$ quoted in this paper. 
Finally, it would be desirable
to estimate the error on the evaluations of $C$, 
$K_{10}$, $K_{11}$ and $K_5+K_6$. In the case of $C$ and $K_{10}$ one can
use experimental data from $\tau$ decay directly into the sum rules and this
should allow for an estimate of the error as well.  
For the others, one must rely on resonance saturation and the reliability of
the estimates can be assessed, in principle, by performing evaluations 
using higher order rational approximants (i.e. including additional resonance
multiplets) and checking the stability of the result.

\noindent{\large\bf 6. Conclusions}

We have discussed an approach to the question of DT violation
(and more generally, to the evaluation of the low-energy constants $K_i$
which parametetrize all electromagnetic effects at order $O(e^2p^2)$)
which is a direct generalization of the classic sum rule of Das et
al.\cite{das1}.  
Instead of the technique of current algebra, as used in \cite{das1}, we have
made use of the chiral lagrangian and the charge spurions. This technique is
rather powerful, as one can judge from the simplicity of the rederivation of
the DGMLY sum rule in sec. 3.
As an immediate consequence
of this method, one finds that all the $K_i$'s can be expressed as a
convolution of a QCD n-point function (with n=2,3 and 4) 
with the photon propagator. We have displayed explicitly the contributions
of the QED counterterms to these constants which allows one, if one wishes,
to perform the calculation using an arbitrary regularization and
renormalization scheme. In the $\overline{MS}$ scheme, the dependence of the
parameters $K_1$ to $K_{13}$ on the gauge parameter $\xi$ and on the $QCD$
renormalization scale $\mu_0$ was entirely determined. 

A current prejudice, based in part on the phenomenology of the usual 
$O(p^4)$ constants $L_i$ (see \cite{egpr}) is that the values of the
constants $K_i$ (and, as a consequence, the amount of DT violation) 
should be essentially controlled by resonance physics at a scale of 1 GeV. 
The great simplicity and the accuracy of the DGMLY expression\rf{dmpi}
certainly make it worthwhile to investigate this prejudice in more detail. 
This assumption concerning the role of the lowest lying resonances
amounts to approximating in a minimal way 
the QCD correlation functions (which occur in 
the convolution expression of the constants $K_i$) 
with rational functions having the
corresponding resonance poles. The parameters of these resonances are then
subject to stringent constraints which can be expressed either in terms of
sum rules that the exact QCD correlation function satisfies or in terms of 
matching conditions at asymptotic momenta. A well known example is the
correlation function $<VV-AA>$ in the chiral limit, which leads to the two
Weinberg sum rules: these can be saturated to a reasonable level of accuracy
by the $\rho$ and the $a_1$ resonances. This model correlator leads to the
DGMLY formula and, as we have shown, to a similar expression for the
constant $K_{13}$. We have displayed a generalization of this construction
to the QCD three-point function $<VAP>$ in the chiral limit. Again, we have
shown that a model containing as poles only those of the $\rho$ and $a_1$
resonances (together with the pion pole) can exactly match all the relevant
asymptotic constraints. 
The properties of the $\rho$ and the $a_1$ resonances are then predicted to
be in reasonable agreement with experiment provided the $a_1$ radiative
width is indeed strongly suppressed as has been suggested by a recent
experiment\cite{condo93}.
These asymptotic constraints cannot be avoided as they ensure
that the QED ultraviolet divergencies are exactly cancelled by the 
QED counterterms. 
We have shown the necessity of not ignoring the $\pi\rho a_1$ couplings (as
was done in earlier work) if
the contribution of the $a_1$ resonance is to be correctly evaluated. 
Under the assumption of dominance of the lowest resonance multiplet in the
vector and the axial-vector sectors, 
the constants $K_{11}$ and $K_{12}$  are  uniquely determined
in terms of $M_\rho$ and $M_{a_1}$.  
The constant $K_{12}$ does not participate
in DT violation but it arises in other interesting isospin violating
phenomena. For instance, as shown in ref.\cite{neufeld}, it is the only
constant $K_i$ which appears in the $Kl_3$ form factor $f^{\kp\piz}_+$ and 
it also appears in the decay constants $F_\pi$ and $F_K$ (note that these
constants are no longer physically meaningful quantities in the presence of
electromagnetism: they are gauge dependent and infrared divergent). These
estimates can in principle be improved by using rational approximants of
higher order, i.e. including more resonances. 

A second class of LEC's that we have considered are $K_9$ and $K_{10}$. We
displayed sum rule expressions in terms of the spectral function differences
$\rho_{V^3}-\rho_{V^8}$ and $\rho_{A^3}-\rho_{A^8}$ or, alternatively, 
$\rho_{V^{ud}}-\rho_{V^{us}}$ and $\rho_{A^{ud}}-\rho_{A^{us}}$ which are
quantities measurable from $\tau$ decay data. In principle,
then, $K_9$ and $K_{10}$ could be evaluated without any recourse to
resonance saturation models using data from $e^+e^-$ scattering as well as
$\tau$ decay. A number of classic chiral sum rules have been already
explored in this manner \cite{d+g}. 
At present, however, there are rather large uncertainties as
far as $\rho_{V^8}$ (or $\rho_{A^8}$) is concerned as is discussed in
ref.\cite{g+k}, but the data is constantly improving. In this sector, the
constant $K_{10}$ is expressed in terms of the double difference 
$\rho_{V^{ud}}-\rho_{V^{us}}-(\rho_{A^{ud}}-\rho_{A^{us}})$ for which simple
resonance approximation can be argued to be a good approximation and yields a
simple expression. Convergence 
is, on the contrary, slower in the case of $K_9$ and one would need several
multiplets of resonances in order to adequately estimate $K_9$. 
Fortunately, as far as DT violation is concerned,
$K_9$ happens to be multiplied by a very small coefficient. Finally, all the
LEC's which are important for DT violation were estimated except for the
combination $K_5+K_6$. The difficulty of estimating $K_5+K_6$ comes from the
complexity of modelling the QCD four-point functions $<AAVV>$ and $<VVVV>$
in terms of resonances. One can think of a number of contributions, for
instance tensor meson exchange, which were never accounted for in DT
violation. We have contented ourselves with a simple order of magnitude
estimate which, after all, could be more reliable than an incomplete 
evaluation.

\noindent{\large\bf Acknowledgments:} I am indebted to Jan Stern for 
pointing out the relevance of the correlators $<VVS>$, $<AAS>$ and $<VAP>$
in DT violation and for many discussions, 
and to Marc Knecht for suggesting the use of the spurion
technique in this context. Jay Watson is thanked for comments on the
manuscript.


\begin{thebibliography}{99}

\bibitem{das1}T. Das, G.S. Guralnik, V.S. Mathur, F.E. Low and J.E.
Young,\prl{18} (1967) 759.

\bibitem{gl85}J.~Gasser and H.~Leutwyler, \np{B250} (1985) 465.

\bibitem{dashen}R. Dashen, \pr{183} (1969) 1245.

\bibitem{kambor}J. Kambor, C. Wiesendanger and D. Wyler, \np{B375} (1996) 266.

\bibitem{anisovich}A.V. Anisovich and H. Leutwyler, \pl{B375} (1996) 335.

\bibitem{socolow}R.H. Socolow, \pr{137} (1965) B1221.

\bibitem{pagels}P. Langaker and H. Pagels, \pr{D8} (1973) 4620.

\bibitem{maltman}K. Maltman and D. Kotchan, \mpl{A5} (1990) 2457.

\bibitem{dhw}J.F. Donoghue, B.R. Holstein and D. Wyler, \pr{D47} (1993) 2089.

\bibitem{bij1}J. Bijnens, \pl{B306} (1993) 343.

\bibitem{b+u} R. Baur and R. Urech, \pr{D53} (1996) 6552.

\bibitem{bij2}J. Bijnens and J. Prades, \np{B490} (1997) 239.

\bibitem{lattice}A. Duncan, E. Eichten and H. Thacker,\prl{76} (1996) 3894. 

\bibitem{urech}R. Urech, \np{B433} (1995) 234.

\bibitem{weinberg}S. Weinberg, Physica {\bf A96} (1979) 327.

\bibitem{gl84}J. Gasser and H. Leutwyler, \ap{158} (1984) 142.

\bibitem{orsay}M. Knecht, B. Moussallam, J. Stern and N.H. Fuchs, \np{B457}
(1995) 513.

\bibitem{berne}J. Bijnens, G. Colangelo, G. Ecker, J. Gasser and M. Sainio,
\pl{B374} (1996) 216. 

\bibitem{dirac}G. Czapek et al., Letter of intent, CERN/SPSLC 92-44.

\bibitem{egpr}G. Ecker, J. Gasser, A. Pich and E. de Rafael, \np{B321}
(1989) 311.

\bibitem{daphne}M. Knecht and J. Stern, in \lq\lq The second Da$\phi$ne
physics handbook", ed. L. Maiani, G. Pancheri and N. Paver, p. 169. 

\bibitem{pdg96}Particle Data Group, R.M. Barnet et al., \pr{D54} (1996) 1.

\bibitem{wrules}S. Weinberg, \prl{18} (1967) 507.

\bibitem{wbook}S. Weinberg, \lq\lq The quantum theory of fields", Cambridge
University Press (1996), vol. 2, sec. 20.5.

\bibitem{peccei}R.D. Peccei and J. Sola, \np{B281} (1987) 1.

\bibitem{d+g}J.F. Donoghue and E. Golowich, \pr{D49} (1994) 1513.

\bibitem{gimenez}V. Gim\'enez, J. Bordes and J. Pe\~narrocha,
\np{B357} (1991) 3

\bibitem{svz}M.A. Shifman, A.I. Vainshtein and V.I. Zakharov, \np{B385}
(1979) 385.

\bibitem{eglpr}G. Ecker, J. Gasser, H. Leutwyler and E. de Rafael, \pl{B223}
(1989) 425.

\bibitem{revecker}G. Ecker, Prog. Part. Nucl. Phys.{\bf 35} (1995) 1.

\bibitem{das2}T. Das, V.S. Mathur and S. Okubo, \prl{19} (1967) 859.

\bibitem{zielinski}M. Zielinski et al., \prl{52} (1984) 1195.

\bibitem{condo93}G.T. Condo et al., \pr{D48} (1993) 3045.

\bibitem{neufeld}H. Neufeld and H. Rupertsberger, \zp {C71} (1996) 131.

\bibitem{tpc} D.A. Bauer et al.,\pr{D50} (1994) R13.

\bibitem{davier} M. Davier, Talk given at the fourth international workshop
on tau lepton physics, Estes Park, Colorado, USA 16-19 September 1996.  

\bibitem{g+k}E. Golowich and J. Kambor, \pr{D53} (1996) 2651.

\bibitem{perez}J.F. Donoghue and A.F. Perez, \pr{D55} (1997) 7075.

\bibitem{b+u2}R. Baur and R. Urech, preprint ZU-TH 30/96, TTP96-54,
hep-ph/9612328.
\end{thebibliography}
\end{document}